\pdfoutput=1
\documentclass[authorversion]{acmart}
\usepackage{colortbl}
\usepackage{times}
\usepackage{float}
\usepackage{longtable}
\usepackage{multirow,multicol}
\usepackage{array,ragged2e}
\newcolumntype{C}[1]{>{\centering\arraybackslash}p{#1}}

\usepackage{xurl} 
\usepackage{hyperref}   
\hypersetup{
    colorlinks=true,
    linkcolor=blue,
    filecolor=blue,      
    urlcolor=black,
    citecolor=blue
    }
\usepackage{booktabs}
\usepackage{multirow}
\usepackage{tikz}
\usepackage{pgfplots}
\usepackage{standalone}
\usepackage{titlesec}
\usepackage{enumerate}
\usepackage{soul}
\usepackage[utf8]{inputenc}
\usepackage{rotating}
\usepackage{afterpage}
\usepackage{multicol}
\usepackage{wrapfig} 
\usepackage{geometry}
\usepackage{caption}
\usepackage{subcaption}

\usepackage{graphicx}
\usepackage{grffile} 

\usepackage[
singlelinecheck=false 
]{caption}
\usepackage[justification=centering]{caption}

\usepackage{subfiles} 

\usepackage{import}
\usepackage{amsmath}
\usepackage{cleveref} 

\usepackage{comment}
\usepackage{ifthen}
\usepackage[colorinlistoftodos,prependcaption]{todonotes} 
\usepackage{xspace} 
\usepackage[normalem]{ulem} 

\usepackage{tikz}
\usepackage{lipsum}

\newcommand\copyrighttext{%
  \footnotesize © {Owner/Author} {2024}. This is the author's version of the work. It is posted here for your personal use. Not for redistribution. }
\newcommand\copyrightnotice{%
\begin{tikzpicture}[remember picture,overlay]
\node[anchor=south,yshift=10pt] at (current page.south) {\fbox{\parbox{\dimexpr\textwidth-\fboxsep-\fboxrule\relax}{\copyrighttext}}};
\end{tikzpicture}%
}
\newboolean{showcomments}
\setboolean{showcomments}{true} 
\ifthenelse{\boolean{showcomments}}
{\newcommand{\nb}[2]{
  \fcolorbox{black}{yellow}{\bfseries\sffamily#1}
  {\sf$\blacktriangleright$\textit{#2}$\blacktriangleleft$}
 }
 
}
{\newcommand{\nb}[2]{}
 
}

\newcommand{\eg}{e.g.,\xspace} 
\newcommand{\ie}{i.e., \xspace} 
\newcommand{\etal}{et. al.\xspace} 

\AtBeginDocument{%
  \providecommand\BibTeX{{%
    \normalfont B\kern-0.5em{\scshape i\kern-0.25em b}\kern-0.8em\TeX}}}

\setcopyright{none}

\begin{document}
\copyrightnotice

\title{Data Mesh: A Systematic Gray Literature Review}

\author{Abel~Goedegebuure}
\email{a.a.goedegebuure@tilburguniversity.edu}
\author{Indika~Kumara}
\email{i.p.k.weerasinghadewage@tilburguniversity.edu}
\author{Stefan~Driessen}
\email{s.w.driessen@tilburguniversity.edu}
\author{Willem-Jan~van~den~Heuvel}
\email{w.j.a.m.vdnHeuvel@tilburguniversity.edu}
\affiliation{%
  \institution{Tilburg University}
  \streetaddress{Warandelaan 2}
  \city{Tilburg}
  \state{North Brabant}
  \country{Netherlands}
  \postcode{5037 AB}
}

\author{Geert~Monsieur}
\email{g.monsieur@tue.nl}
\author{Damian~Andrew~Tamburri}
\email{d.a.tamburri@tue.nl}
\affiliation{%
  \institution{Eindhoven University of Technology}
  \streetaddress{Groene Loper 3}
  \city{Eindhoven}
  \state{North Brabant}
  \country{Netherlands}
  \postcode{5612 AZ}
}

\author{Dario~Di~Nucci}
\email{ddinucci@unisa.it}
\affiliation{%
  \institution{University of Salerno}
  \streetaddress{Via Giovanni Paolo II, 132}
  \city{Fisciano SA}
  \state{Salerno}
  \country{Italy}
  \postcode{84084}
}

\renewcommand{\shortauthors}{Abel and Indika, et al.}

\begin{abstract}
Data mesh is an emerging domain-driven decentralized data architecture that aims to minimize or avoid operational bottlenecks associated with centralized, monolithic data architectures in enterprises. The topic has picked the practitioners' interest, and there is considerable gray literature on it. At the same time, we observe a lack of academic attempts at defining and building upon the concept. Hence, in this article, we aim to start from the foundations and characterize the data mesh architecture regarding its design principles, architectural components, capabilities, and organizational roles. We systematically collected, analyzed, and synthesized 114 industrial gray literature articles. The review provides insights into practitioners' perspectives on the four key principles of data mesh: data as a product, domain ownership of data, self-serve data platform, and federated computational governance. Moreover, due to the comparability of data mesh and SOA (service-oriented architecture), we mapped the findings from the gray literature into the reference architectures from the SOA academic literature to create the reference architectures for describing three key dimensions of data mesh: organization of capabilities and roles, development, and runtime. Finally, we discuss open research issues in data mesh, partially based on the findings from the gray literature.   
\end{abstract}



\maketitle

\section{Introduction}
The world is living in the golden age of data. IDC (International Data Corporation) predicts that by 2025, the amount of digital data generated by enterprises and individuals will grow 61\% to 175 zettabytes~\cite{rydning2018digitization}. While the advances in the analytic data landscape can enable organizations to turn their ever-growing raw data into value, the difficulties in the integration, management, and governance of data at scale hamper the success of data analytic projects of organizations~\cite{deloittedm,JANSSEN2020101493}. In the current practices of centralized data management, domain teams rely on a central data management team to collect, process, and manage domain data. This central team is becoming the bottleneck for unlocking the value from domain data. There is a need to shift the domain data's responsibility from the central team to domain teams~\cite{dehghani2022data}.

\begin{figure}[h!]
    \centering
    \includegraphics[scale=0.5]{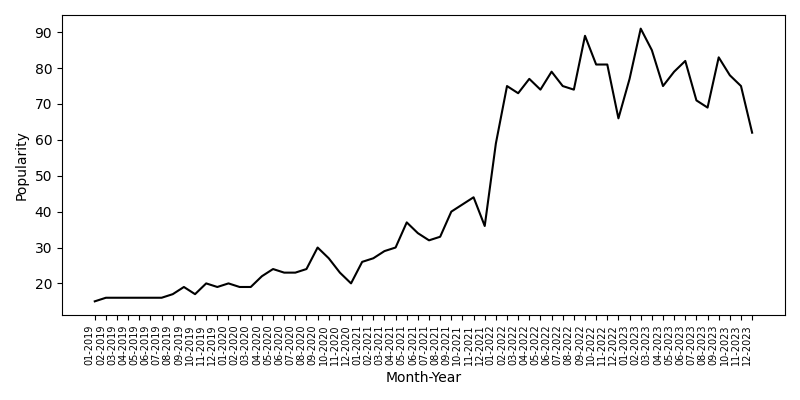}
    \Description{Google trends for the term ``Data Mesh".}
    \caption{Google trends for the term ``Data Mesh".}
    \label{fig:googl}
\end{figure}

A data mesh is emerging as a novel decentralized approach for managing data at scale by applying domain-oriented, self-serve design and product thinking~\cite{dehghani2022data}. Zhamak Dehghani first defined the term data mesh in 2019~\cite{S2}. \Cref{fig:googl} shows the search index for ``Data Mesh" on Google Trends from January 2019 to December 2023 (monthly average). The interest in data mesh grew gradually until early 2022. Then, there was a sharp increase, followed by a generally sustained popularity. Despite its popularity, little academic literature exists on the data mesh topic. However, as organizations increasingly investigate data mesh, we saw an opportunity to systematically review the gray literature to better characterize the concept and drive a research agenda. 

Zhamak~\cite{dehghani2022data} defines the architecture of data mesh based on four principles: data as a product, domain ownership of data, self-serve data platform, and federated computational governance. The data assets in an organization are offered as products that enable consumers to efficiently and autonomously discover and use domain data. Domain teams closest to the origin of data create and manage data products. A mesh of data products emerges when domains consume data products from each other. To ensure the data products are interoperable and compliant with internal and external regulations and policies, a federated governance team is responsible for defining and enforcing product interoperability standards and compliance policies across the data mesh. Finally, a self-serve data platform supports the domains by offering infrastructure and platform services necessary to build, maintain, and manage data products. 

To a certain extent, we observed a strong analogy between data mesh and service-oriented architecture (SOA), particularly with domain-driven microservices~\cite{dehghani2022data,S2,S110}. SOA represents a software application as a set of loosely coupled collaborating services~\cite{papazoglou2007service}. It promotes the idea of building applications by composing existing applications from multiple parties through discovering and invoking network-available services. The microservices architecture is a variant of SOA. The success of the domain-oriented design and microservices architecture was an inspiration for inventing the data mesh architecture~\cite{S1,S2,dehghani2022data}. Data products are often exposed as services and consumed through APIs. A data mesh relates and connects data products, which is analogous to a service mesh. The product owners can publish their data products, and consumers can discover and interconnect the desired products to perform cross-domain data analysis and build new value-added data products, which is analogous to service composition. 

In this paper, we report a Systematic Gray Literature Review (SGLR) that aims to identify and review the gray literature on the data mesh to synthesize the practitioners' understanding of the data mesh and identify research challenges for academia. We selected and analyzed 114 gray literature articles. Our findings include the practitioners' characterization of four principles of a data mesh, common architectural components in a data mesh, and potential benefits and drawbacks of a data mesh. Moreover, the similarities between SOA and data mesh motivated us to adopt the well-established reference architectures from SOA to organize the findings from our SGLR into three reference architectures. Our mapping of data mesh to SOA can also help the research community understand research challenges in data mesh and the potential use of SOA research results to address those challenges. 

The remainder of this article is organized as follows. \Cref{methodology} describes our systematic literature review methodology. Based on the findings of the gray literature, \Cref{definition} explores the data mesh and its four principles and \Cref{benefits} presents its potential benefits and drawbacks. \Cref{referencearchitecture} describes our three reference architectures for the data mesh, while \Cref{research-agenda} lists potential research challenges. \Cref{background} summarizes the academic research studies on data mesh and the literature review studies relevant to our survey. Finally, we discuss the threats to the validity of our research in \Cref{discussion} and conclude the paper in \Cref{conclusion}. 
\section{Methodology}
\label{methodology}
\begin{figure}[ht]
 \centering
 \includegraphics[width=0.8\textwidth]{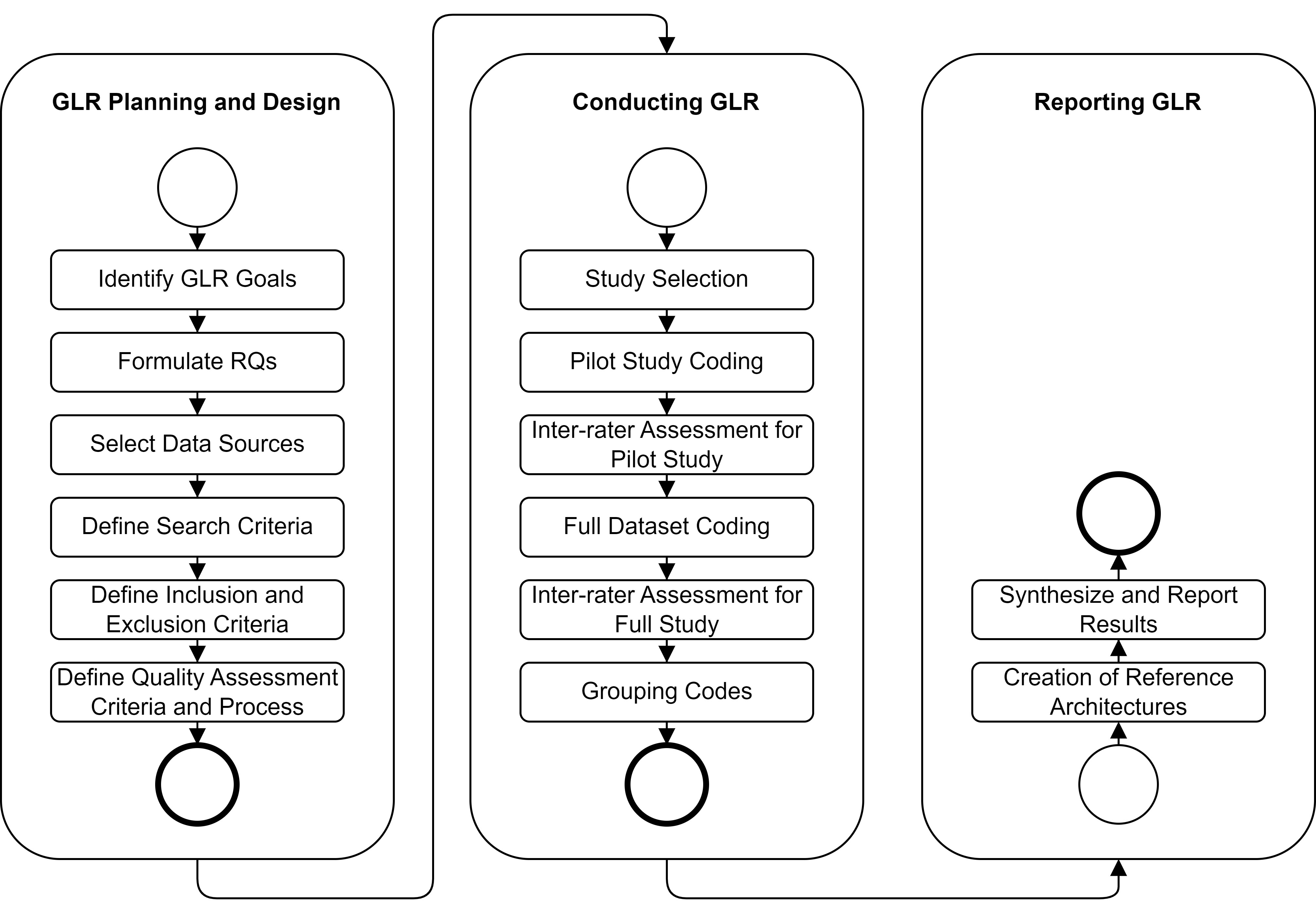}
 \Description{Systematic gray literature review process.}
 \caption{Systematic gray literature review process.}
 \label{fig:GLR}
\end{figure}

Due to the novelty of the data mesh paradigm, little academic literature is available on the topic. At the same time, since the data mesh concept has its roots in the industry as opposed to academics~\cite{S1,S2}, gray literature about the topic, such as blog posts and white papers, is widely available. Since the quality of structured literature reviews (SLRs) depends greatly on the availability of enough source material, a gray literature review (GLR) was conducted to investigate the state-of-the-art for the data mesh. In this work, we follow the guidelines provided by Garousi \etal ~\cite{Garousi2019}, who adapted the SLR guidelines of Kitchenham and Charters~\cite{Kitchenham2007} specifically for GLRs. Following these methodologies, our GLR is organized into three stages: a planning and design phase, an execution phase, and a reporting phase. A high-level overview of these phases and their steps are depicted in \Cref{fig:GLR}.

\subsection{Planning and Design of the Gray Literature Review} Our main goals were to provide a unified definition of the data mesh concept, create reference architectures for the data mesh, and identify research challenges. We formulated four research questions to achieve these goals:

\begin{description}
 \item[RQ1. What is a data mesh?] A data mesh embodies four design principles~\cite{S1,S2,dehghani2022data}: data as a product, domain-driven data ownership, self-serve data platform, and federated computational governance. Practitioners may have different perspectives on these principles and experiences applying them. Hence, RQ1 aims to understand and characterize the four principles of data mesh from a practitioner's standpoint. 
 
 \item[RQ2. Why is the data mesh needed?] Specifically, what are the benefits and concerns of adopting it, and when should it be used? A data mesh has advantages and disadvantages and is not a one-size-fits-all approach for data management. Thus, RQ2 aims to understand the data management issues that data mesh tries to address and the potential benefits, use cases, implementation challenges, and drawbacks of data mesh.
 
 \item[RQ3. How should an organization build data mesh?]
 Some organizations have started their data mesh journey and report their experience, including development and operation guidelines, options, and examples. Hence, RQ3 aims to analyze the relevant gray articles to create a set of reference architectures to guide organizations in making better design choices when embarking on their data mesh journey. Furthermore, due to the novelty of the data mesh architectures, we aim to build reference architectures by mapping findings from the gray literature into the architectures used in comparable domains. 
 
 \item[RQ4. What are the research challenges concerning data mesh?]
 Practitioners report critical challenges in the gray articles; hence, RQ4 aims to identify research challenges concerning data mesh based on practitioners' challenges. 
\end{description}

\begin{figure}[ht]
 \centering
 \includegraphics[width=\textwidth]{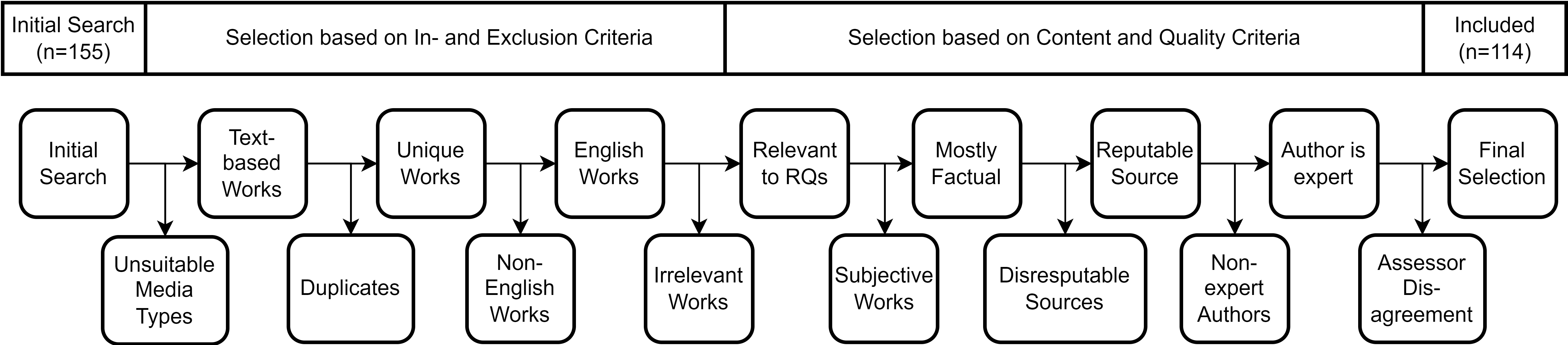}
 \Description{The search and selection process. Steps are shown sequentially, and each step reduces the number of candidate sources until the final selection. After the final exclusion step, an inter-rater reliability test---featuring the well-established Cohen Kappa coefficient calculation---was performed with $\kappa = 0.79$ to ensure that reviewer bias was not inappropriately high.}
 \caption{The search and selection process. Steps are shown sequentially, and each step reduces the number of candidate sources until the final selection. After the final exclusion step, an inter-rater reliability test---featuring the well-established Cohen Kappa coefficient calculation---was performed with $\kappa = 0.79$ to ensure that reviewer bias was not inappropriately high.}
 \label{fig:search_and_selection}
\end{figure}

Following the protocols by Kitchenham \etal~\cite{Kitchenham2007}, we defined several search strings to identify relevant gray literature sources. Then, each string was applied to our selected search engine (Google), after which our pre-established inclusion and exclusion criteria were used to refine the results as shown in \Cref{fig:search_and_selection}.

As is common practice in GLRs, our search string selection process was based on trial queries with different search strings, which can indicate the usefulness of that search string~\cite{Straus2014}. We started with the keywords used in our research questions. We kept improving the combination until we settled on the following query, which yielded promising results for answering our research questions.

\begin{description}
	\item[Query:] Data Mesh (Requirements \textbar~Benefits~\textbar~ Drawbacks ~\textbar~Concerns \textbar~Advantages~\textbar~Disadvantages \textbar~Challenges~\textbar~Components \textbar~Architecture \textbar~Design)
\end{description}

The query was then executed in the Google search engine on September $1^{st}$ 2021, and all search results were checked against our inclusion and exclusion criteria. 
Due to the relative novelty of data mesh, information and sources at that time were mostly limited to the generic data mesh principles. Therefore, the GLR was extended to include data sources about the architectural design of data mesh until May $1^{st}$ 2022. This step resulted in 155 data sources from 2019 until 2022.

To effectively process our sources in line with grounded theory, we limited our gray literature source selection to text-based sources, such as reports, blog posts, white papers, official documentation from vendors, presentations, and transcriptions of keynotes and webinars. Additionally, duplicate and non-English sources were eliminated to remove bias and ensure the text was legible to the authors.

To guarantee the relevance and quality of the sources used for this GLR, we applied several quality criteria inspired by previous gray literature reviews \cite{Garousi2019,Butijn2020,kumara2021s}. As a first step, we eliminated any work that does not explicitly contain either a theoretical discussion or a practical implementation of a data mesh that can aid us in answering our proposed research questions. Examples of the former include a formal definition, an example architecture, or an overview of the benefits and challenges of data mesh implementation. Examples of the latter include discussing tools or an instantiation of (components of) a data mesh.

Next, works that consist primarily of subjective statements that promote the author's opinion, as opposed to factual information, were excluded. Furthermore, we checked whether the publishing organization is respected in information technology and whether the author is a reputable individual, \ie they are associated with a reputable organization, have reputable expertise, or have published other articles in the field of data management.

Finally, two authors checked each article against the content and quality criteria to eliminate potential selection bias. A mutual agreement was required for final inclusion in the data set, and disagreements were discussed between the authors. The resulting inter-rater reliability assessment~\cite{saal1980rating} yielded a Cohen Kappa of 0.79, which indicates substantial agreement between the authors.

\begin{figure}[ht]
 \centering
 \begin{subfigure}[b]{0.49\textwidth}
 \centering
 \includegraphics[width=\textwidth]{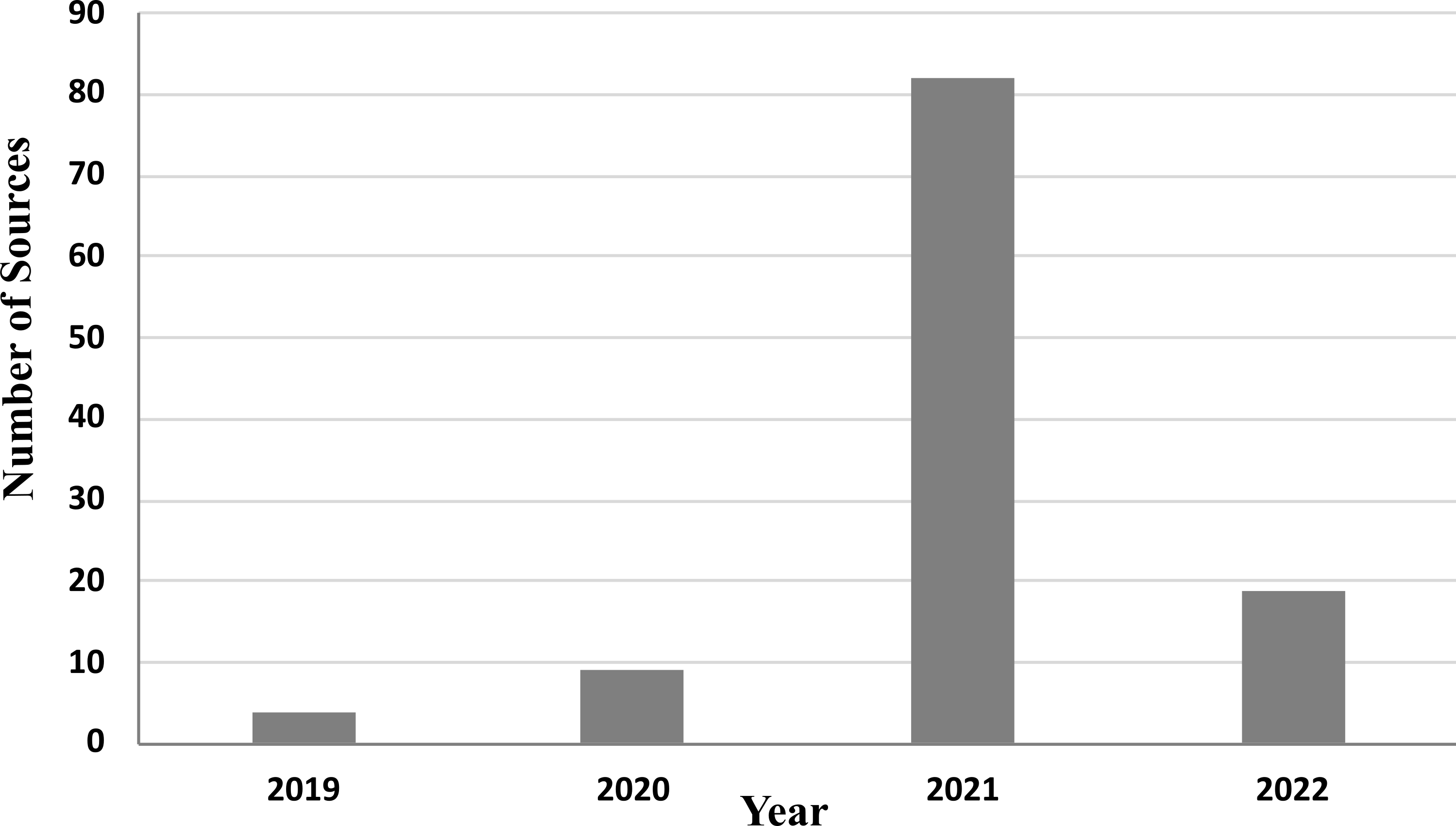}
 \caption{The publication date of the selected sources. }
 \label{fig:year_dist}
 \end{subfigure}%
 \hfill
 \begin{subfigure}[b]{0.49\textwidth}
 \centering
 \includegraphics[width=\textwidth]{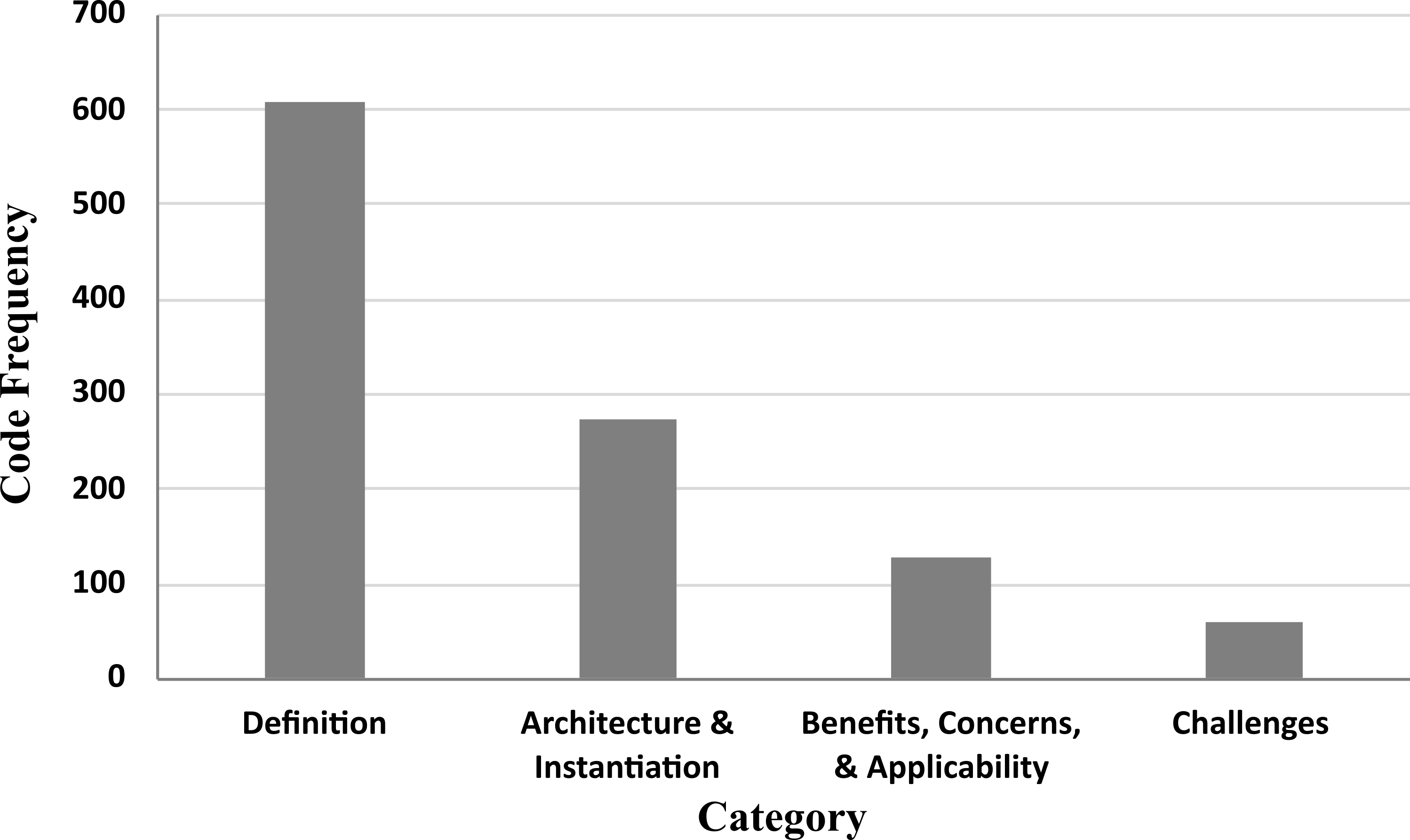}
 \caption{Topics discussed in the analyzed studies.}
 \label{fig:categories}
 \end{subfigure}%
 \Description{Source statistics}
 \caption{Source statistics.}
 \label{fig:source-stas}
\end{figure}

The final selection consists of 114. \Cref{fig:year_dist} shows the distribution of the selected sources. A full overview of each source can be found in our online appendix (see~\Cref{replicaion-package}).

\subsection{Conducting and Reporting the Gray Literature Review}
\label{sec:conducting_the_GLR}
The second set of steps is related to the execution of the review in line with a qualitative analysis methodology. Structural coding was used to get an initial overview of the topics, after which descriptive coding was used to summarize parts of the topics. Structural coding was used to identify larger text segments on a broad topic. Descriptive coding was then used to go more in-depth on these topics by summarizing the basic topic into a word or short phrase~\cite{saldana2021coding}. We leveraged Atlas.ti\footnote{\url{https://atlasti.com/}} for the coding process.

Like~\cite{kumara2021s}, an initial set of 20 sources was randomly selected and analyzed to establish an initial set of codes. These were discussed between the first three authors of the paper. The entire dataset was coded once consensus on the initial set of codes was reached. The grouping of codes into topics and the identification of categories were established through discussion by the first three authors of this paper.

We first used the data extracted from the gray literature using the established codes to answer RQ1 and RQ2. Next, to address RQ3, we created three reference architectures by mapping our findings to the reference architectures used in service-oriented computing~\cite{papazoglou2008web,Papazoglou2,servicemesh,SDSN}. The first three authors of the paper collaboratively created the reference architectures over a set of meetings. The other authors validated such architectures. We addressed any discrepancies through discussions. We followed a similar approach to answer RQ4. We identified the potential research challenges by mapping the difficulties mentioned by the practitioners in the gray articles to the relevant research issues from the service-oriented computing~\cite{Papazoglou2,WeiRe,papazoglou2007service} and data management~\cite{dustdar2012quality,stonebraker2018data,al2019systematic,JANSSEN2020101493}.

The results from the systematic gray literature are presented in the rest of this paper. We defined four main categories as shown in \Cref{fig:categories}. In the online appendix, we provide the details of each category, including topics, underlying atomic codes, and the sources that address these codes. 


\subsection{Replication Package}
\label{replicaion-package}
We made the related data available online\footnote{https://github.com/IndikaKuma/DataMesh} to enable further validation and replication of our study.
It contains the full list of sources, the qualitative analysis (i.e., codes, groupings, and analysis) performed with the \textit{Atlas.ti} tool, and code tables.
\section{RQ1. What is Data Mesh}
\label{definition}

In short, a data mesh is a domain-oriented decentralized architecture for managing (analytical) data at scale. It enables the decomposition of an organization's monolithic analytical data space into data domains aligned with business domains. Such decomposition moves the responsibility of managing and providing high-quality data and valuable insights from the conventional central data team into domain teams that intimately know the data. 

Zhamak~\cite{S1,S2,dehghani2022data} defines the data mesh concept using four principles: (i) data product thinking, (ii) domain ownership of data, (iii) federated computational governance, (iv) and self-serve data platform. Based on the findings from our gray literature, \Cref{dataproducts,domainownership,governance,self-serve} aim to study the practitioners' perspective on these principles and how they expand or implement the principles.
 
\subsection{Principle 1: Data as a Product} \label{dataproducts}

This principle applies product thinking to analytical data, offering it as a valuable asset to potential data consumers. 

\paragraph{Elements of a Data Product.} Zhamak~\cite{S1,S2,dehghani2022data} defines a data product as an \textit{architectural quantum} (an independently deployable, high-cohesive component encompassing all the structural elements required for its function~\cite{ford2022building}). A data product is a composition of all its key elements: 1) data and metadata, 2) code for ingesting, processing, serving, and governing data, and 3) infrastructure for building, deploying, and executing code and storing data and metadata. The metadata of a data product may include information such as product owners, access endpoints,  data models, access policies, data quality metrics, and data lineage. A data product is a node in the data mesh and an autonomous entry that receives or takes the input data from one or more sources, applies some computation/transformation on the received data, and produces and serves the results expected by the consumers. 

While most gray literature generally agrees with Zhamak's definition of a data product as an architectural quantum, some sources~\cite{S13,S24} consider a curated data set (without code and infrastructure) a data product. Several sources also introduced the concept of \textit{
a data contract}~\cite{S54, S60, S73, S110, S108, S113} to represent the interface of a data product. It can include data access endpoints and the syntax, semantics, quality, and terms of use of the data offered by the data product. The practitioners proposed implementing data contracts by storing the relevant metadata or defining them as API using standards such as Open API~\footnote{https://www.openapis.org/}. 

Zhamak recommends using the infrastructure as code (IaC) approach to automate the provisioning of the infrastructure of a data product~\cite{dehghani2022data}. The gray literature also highlighted the importance of IaC~\cite{S8, S37, S97, S98, S108} and provided several examples of data products that use IaC for deploying them~\cite{S13,S45,S73,S75}. IaC employs the source code to manage and provision infrastructure resources and deploy and configure applications, enabling the application of software engineering tools and best practices~\cite{IaC2020}.

\paragraph{Types of Data Products.}

Zhamak~\cite{S1,S2,dehghani2022data} mentioned three types of data products: source-aligned, aggregates, and consumer-aligned. The gray literature also uses the same terminology for categorizing data products. Source-aligned data products~\cite{S6,S20,S87,S88,S99,S106,S110,S112} ingest and process data from source systems such as operational databases, external APIs, and sensors. The processed data is then offered as a product to be consumed by the organization's end-users or downstream data products. The aggregate data products~\cite{S38,S86,S110} compose the data from the upstream data products to produce the value-added data for downstream consumers. Compared to the aggregates, consumer-aligned data products~\cite{S6,S87,S88,S95,S96,S98,S109,S110,S112} are generally not sold as data products to other domains; instead, they serve data to implement the end-user use cases of the organization.

\begin{figure}[ht!]
 \centering
 \includegraphics[scale=0.8]{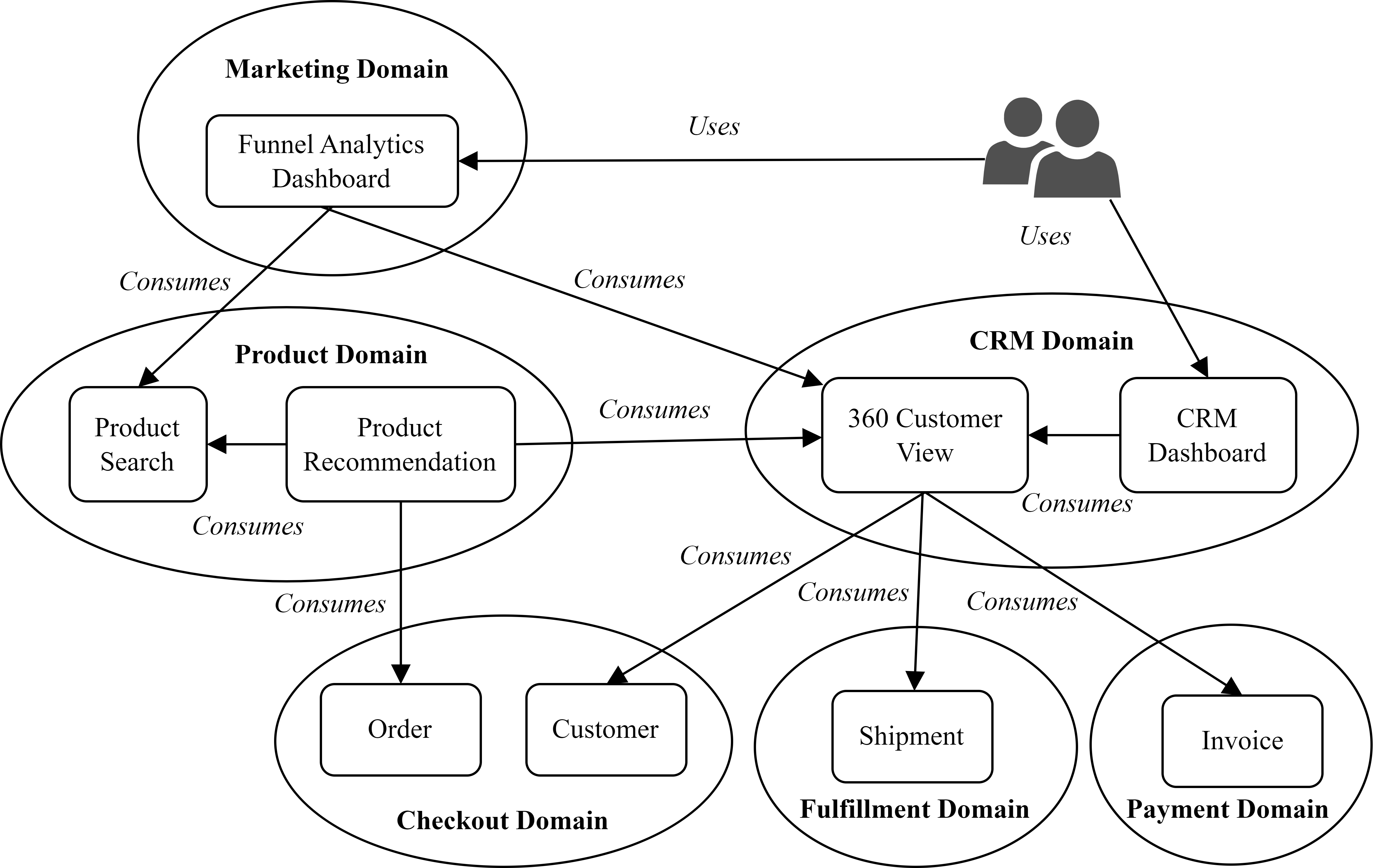}
 \Description{An example of data products in different domains, adopted from~\cite{S110}.}
 \caption{An example of data products in different domains, adopted from~\cite{S110}.}
 \label{fig:dataproduct-example}
\end{figure}

\Cref{fig:dataproduct-example} shows data products spread across domains. Some data products, \eg product search, order, and invoice, are source-aligned products. They mostly ingest raw data from the underlying operational data sources (\eg product and order databases), clean and transform the raw data, and store them locally. They also provide the interfaces to serve their local domain data to other data domains. The aggregate data products, \eg 360-degree customer views and product recommendations, aggregate data from one or more domains and offer as a value-added product to one or more domains. Some data products, \eg funnel analytics and CRM dashboards, primarily serve end-users (\ie consumer-aligned data products).

\paragraph{Characteristics of Data Products.}
Zhamak initially introduced six properties for a data product~\cite{S2} and later added two properties~\cite{dehghani2022data}. From the gray literature, we tried to identify the practitioner's perspectives on them.

\begin{description}
 \item[Discoverable.] Data consumers should be able to discover the data products in an organization easily. A common approach proposed in the gray literature is to use a central data/product catalog to store the metadata of data products~\cite{S4, S6, S7, S8, S10, S11, S13, S14, S15, S17, S28, S29, S32, S33, S35, S39, S40, S45, S47, S49, S57, S58, S63, S64, S73, S78, S79, S80, S81, S83, S82, S84, S88, S89, S98, S101, S103, S104, S107, S110, S111}. The data catalog can enable consumers to search for data products using metadata. However, Zhamak~\cite{dehghani2022data} does not recommend a central data catalog managed by a separate team, as data products are responsible for deciding and providing their discoverability information. Instead, the data product can offer a discovery interface that can be used to build product search and exploration capabilities at the mesh level.

 \item [Interoperable.] It should be (relatively) easy to combine or integrate data from different data products into composite data products. The consistent use of open standards and conventions across data products can increase interoperability between data products. The gray literature mentions several standards and practices a data mesh architecture should encompass~\cite{S9,S15, S21,S27,S32,S47,S49,S82,S83,S98,S101, S110}. For example, the open standards for describing data products and their composition, including interfaces, metadata, and (access and management) policies, can ensure the interoperability of data products within and across organizations. In addition, standard metadata fields and business glossary models can guarantee that all data products are described unambiguously. Finally, a common method for data quality modeling can ensure all domains define their products' data quality concerning the same KPIs (key performance indicators).
 
 \item[Addressable.] The interfaces of a data product should provide unique, programmatically accessible endpoints. The interfaces include those for ingesting and serving data and discovering and managing data products~\cite{dehghani2022data}. The gray literature primarily focused on data access endpoints and mentioned examples such as web service APIs~\cite{S15,S24,S29,S30,S71,S78,S98,S108,S113}, storage buckets~\cite{S6, S13, S14, S29, S42, S51, S53, S69, S89, S98, S104, S107}, topics in publish-subscribe messaging systems~\cite{S6, S15, S24, S29, S38, S42, S45, S55, S67, S69, S71, S73, S78, S82, S98, S100, S110}, and SQL endpoints~\cite{S16,S24,S102}.

\item[Natively Accessible.] Zhamak~\cite{dehghani2022data} emphasizes that a data product should provide multiple endpoints to consume its data so consumers can use their preferred methods to access data products. Along the same lines, the gray literature mentioned that the data product needs to support polyglot data output interfaces~\cite{S7, S29, S53, S64, S78, S97, S106}. 
 
\item[Understandable.] The data models and natural language documentation of data products should help potential consumers understand the products' semantics, syntactic, and behaviors/usage~\cite{dehghani2022data}. For example,  the documentation of a data product can include the entities in the data the product consumes and produces, the relationship between those entities, the relationships with the other data products in the mesh, and the information about access patterns and how to request the access to data~\cite{S9,S15,S55,S59,S66,S97,S99}. The information about product owners and subject matter experts can ensure that the consumers know whom to contact when needing more information related to the data. The schemas for all the data views can be used to describe the data structure. Service level objectives (SLOs) can describe the product's quality and act as a data contract. The metadata of data lineage and provenance can show the origin of the data and give consumers trust in the data product as it becomes clear how the data is created and processed. The literature suggests using common data modeling languages or standards can help improve the understanding of data/messages exchanged between data products~\cite{S15,S55}.
 
\item[Secure.] A data product must enforce the necessary security policies to ensure the consumers can use it in a secure and regulatory-compliant way~\cite{dehghani2022data}. The policies include but are not limited to data access control, data encryption at rest and data in transit, data confidentiality, and data retention. In the data mesh architecture, while the security policies are primarily managed centrally, their enforcement happens in a decentralized fashion~\cite{dehghani2022data}. The \textit{policy-as-code} approach, where policies are defined and enforced using code in a high-level language, was recommended by Zhamak~\cite{dehghani2022data} and other practitioners~\cite{S21,S29,S54,S78,S110,S113} for managing security policies of data products. 
 
\item[Trustworthy.] Data consumers should be able to trust the data coming from a data product. The gray literature provides several guidelines to enhance the trustworthiness of data products. Product owners can use a methodology provided by the federated governance team to model data quality to define their own product's quality on several dimensions (e.g., completeness, accuracy, and timeliness)~\cite{S3, S13, S58, S108}. Moreover, the self-serve platform can support product developers in ensuring quality throughout the product's lifetime (\eg automated monitoring of data quality issues and execution of corrective actions)~\cite{S55,S72}. Product owners and consumers can also use a formal contract that defines a set of service-level objectives (SLOs), reflecting the target quality for each data product ~\cite{S64,S79,S99,S113}. Finally, another approach to improve the trustworthiness of data is to provide data provenance and lineage information to the product consumers so that they can inspect the origin and history of the data they use~\cite{S15,S53,S88}.

\item[Valuable on its own.] The data offered by a product should possess long-term value to the rest of the organization. Even without being aggregated with other data products, a data product should support meaningful use cases~\cite{dehghani2022data}. However, the gray literature mostly provides the use cases where the data from one domain becomes valuable to the organization once it is combined or correlated with the data from other domains, for example, user behavior data and sales information~\cite{S6} and smart device data and enterprise data~\cite{S108}. The value of a data product should be continuously monitored and maintained. The gray literature mentioned adopting the DataOps methodology~\cite{DataOpsDef} for automating the continuous delivery of high-quality data products to consumers~\cite{S21,S37,S78,S98}. 
\end{description}

\subsection{Principle 2: Domain Ownership of Data} \label{domainownership}
A central data team often manages and shares analytical data in traditional data architectures. In contrast, data mesh distributes the ownership of data to the teams or individuals (belonging to the domains in the organizations) closest to the data and knows the data intimately~\cite{dehghani2022data}. Autonomous domain teams are responsible for building and managing data products from domain data. Zhamak~\cite{S1,S2,dehghani2022data} advocates applying the Domain-Driven Design (DDD) approach~\cite{evans2004domain} to decompose the analytical data in an organization into data domains that align with organizational units. 

\paragraph{Domain Formation.} 
The gray literature is in agreement with Zhamak regarding the use of DDD for identifying data domains and their data products~\cite{S29, S35, S45, S46, S55, S59, S75, S78, S86, S87, S95, S96, S97, S98, S103, S110, S112}. DDD decomposes an organization's business functions and capabilities into domains, enabling the designing of software systems based on models of the underlying domains~\cite{evans2004domain}. A domain model describes selected aspects of a domain using the vocabulary (\ie abstractions) shared by domain experts, users, and developers. Organizations already use separation of responsibilities by decomposing the organization into different business domains (\eg products, sales, and marketing). 
Data products can be built around operational and analytical data produced by these business domains. Hence, business domains form logical constituents for the distribution of data ownership and may be decomposed into subdomains to manage their complexity better. \Cref{la-ra} discusses the application of DDD for building data products. 

\paragraph{Responsibilities.}
Zhamak~\cite{S1,S2,dehghani2022data} identified two key responsibilities for domain teams: building and maintaining data products and carrying out governance activities at the domain level. The gray literature also mentions similar responsibilities~\cite{S4,S9,S13,S16,S20,S21,S22,S24,S25,S29,S33,S35,S41,S42,S43,
S47,S49,S51,S53,S58,S63,S64,S67,S71,S73,S74,S77,S88,S87}. Domain teams possess in-depth knowledge about domain data, including potential use cases and consumers of data, as well as the processes and techniques for maintaining data quality. Hence, they can determine which domain data assets should be provisioned as data products. They can use the tools and services the self-serve data platform provides to create, manage, and serve data products (see \Cref{self-serve}). Domain teams also carry out governance activities to ensure compliance and conformance of data products to relevant regulations, organizational policies, and standards (see \Cref{governance}). 

\subsection{Principle 3: Federated Computational Governance} \label{governance}
Data mesh uses a decentralized governance model in which sovereign domains govern their data products~\cite{S1,S2,dehghani2022data}. Overarching governance on a global level across the mesh of the data products mainly aims to ensure interoperability between data products and to enforce standardizations and organization-wide policies~\cite{S3,S4,S9,S15,S18,S19,S21,S25,S33,S37,S38,S53,S58,S60,S62,S63,S74,S77,S82,S95,S111}. Balancing the levels of centralization and decentralization can be challenging in the federated governance model. On the one hand, centralization gives more control to ensure interoperability and compliance. On the other hand, the domains need sufficient autonomy to build data products that align with their business. 

\paragraph{Global Governance.} We refer to the governance activities conducted by the federated governance team on the data mesh level as global governance. The federated governance team comprises different kinds of representatives such as domain experts, subject matter experts, governance experts, members from the legal department, the platform team, and the management~\cite{dehghani2022data}. The common global governance activities include:

\begin{description}
 \item[Defining and Enforcing Organization-wide Standards.] A key goal of standards is to increase the interoperability of data products~\cite{S1,S2,dehghani2022data}. Standards can also used to ensure data products adhere to security and privacy requirements modeled by the federated governance team. Zhamak~\cite{dehghani2022data} emphasized the standardization of the interfaces of data products, data and query modeling, and data lineage. The gray literature proposed several standards: W3C data catalog vocabulary for product metadata modeling~\cite{S49}, cloud events for event data~\cite{S20}, Open API for product interfaces~\cite{S30,S110}, JSON schema for data modeling~\cite{S21,S110}, and ONNX (Open Neural Network Exchange) for machine leaving models~\cite{S110}.     
  
 \item[Defining and Enforcing Data Governance Policies.] Data governance policies define rules or guidelines that enforce consistent and correct collection, storage, access, usage, and management of data assets in an organization. Zhamak~\cite{dehghani2022data} recommends cross-cutting concerns such as regulatory compliance, user identity management, and product ownership management for mesh-level governance policies. Several gray literature articles presented examples of using a central identity and access management (IAM) provider to manage users and their permissions~\cite{S13,S103} and a data catalog for automatically discovering and classifying sensitive data~\cite{S13,S73,S103}.
  
 \item[Developing a Methodology to Define and Assess Data Quality.] The federated governance team is also responsible for creating a methodology to define data quality to be enforced by data product owners~\cite{dehghani2022data}. \textcolor{black}{According to literature sources~\cite{S3,S21,S55,S58,S11}, a standardized methodology and tools can ensure that all data products define quality on a common set of quality dimensions with shared definitions. A lack of consistent data quality enforcement will lead to data swamps~\cite{S35,S45}. The practitioners propose using a common format for data quality constraints and results~\cite{S25} and metadata rules for enforcing the consistent modeling of concepts  shared across domains~\cite{S64}.}

 \item[Maintaining a Common Business Glossary.] Organizations use business glossaries of business terms and their definitions to ensure the same definitions are used throughout the organization when analyzing data. The data mesh principles discourage creating one canonical data model for the whole organization, which goes against decentralization~\cite{dehghani2022data}. \textcolor{black}{However, some gray literature sources recognize the importance of having a shared definition of concepts used throughout the organization (\eg master and reference data such as customers and currency codes) as this contributes to the interoperability of data products~\cite{S38,S90,S111,S114}.} 
  
 \item[Monitoring Data Mesh.] \textcolor{black}{The federated governance team should continuously observe and assess the overall health of the data mesh and make necessary interventions. Zhamak~\cite{S1,S2,dehghani2022data} discusses collecting metrics such as lead time to change a data product, the number of data product consumers, and data quality dimensions. They can be used to determine the value of a data product and make interventions such as promoting or demoting products. The gray literature sources also emphasized the importance of defining and measuring KPIs (Key Performance Indicators) such as levels of interoperability, compliance, and usage of data products~\cite{S4,S48,S71,S88,S110}. }
 
\item[Creating Incentive Models.] Provisioning data products will add to the domain teams' normal workload and will likely not fit their traditional job descriptions~\cite{S21}. For example, a supply chain department, whose main task is to optimize delivery routes, may not be keen on providing high-quality data to other domains. Without proper incentives, domains might not see a reason to build data products. Thus, organizations implementing or transforming to a data mesh architecture should develop an incentive model to motivate domains to provision their data as products and continuously govern the offered products~\cite{S1,S2,dehghani2022data}. \textcolor{black}{Several gray literature articles emphasized the need for aligning the incentives of data producers and data consumers and motivating data producers to generate the operational metadata necessary to assess the quality and trustworthiness of data products~\cite{S1,S2,S13,S16,S24,S37,S55}.} 

\end{description}

\paragraph{Local Governance.} We refer to the governance activities conducted by the product/domain teams on a product/domain level as local governance. The product owner, potentially in combination with other domain members, is responsible for conducting the local governance activities. \textcolor{black}{Zhamak~\cite{S1,S2,dehghani2022data} identified several governance activities at the domain level: domain data modeling, data quality assurance, and execution of data access control policies.}  The gray literature mentions similar responsibilities for the local governance team:
\begin{description}
 \item[Managing the Data Models of the Product.] \textcolor{black}{The product team possesses domain knowledge and thus can accurately model the product's data~\cite{S13,S16,S24,S37,S55,S64}. The schemas of such models need to be evolved accordingly to reflect the changes in the domain's data sources~\cite{S13,S72}. }
 \item[Managing Data Access.] \textcolor{black}{Domain experts thoroughly understand the data and can, therefore, best judge who should and should not have access to (which parts of) the data~\cite{S29,S64,S86,S98}. In addition, tasks such as collecting data lineage, implementing data retention policies,  and encrypting data can also be implemented at the product level~\cite{S13,S28,S98}. } 
 \item[Managing Compliance and Conformance.] Like managing access controls, ensuring compliance and conformance throughout the product's lifetime requires a thorough understanding of the domain data~\cite{dehghani2022data,S28,S86}. \textcolor{black}{The examples mentioned by the gray literature include the adherence to the data modeling standards~\cite{S63,S74} and interoperability standards~\cite{S80,S11}, and ensuring privacy compliance~\cite{S28,S98}.}
 \item[Managing Data Quality.] Using the quality assessment methodology defined by the federated governance team, the product team establishes the product's target quality based on a set of quality dimensions~\cite{dehghani2022data}. \textcolor{black}{The data quality tests mentioned in the gray literature include measuring data quality dimensions, validating schemas, and checking for violations of domain-specific business rules~\cite{S55,S57,S63,S64,S88}.}
 \item[Monitoring Data Product Health] Throughout the lifetime of the product, the product owner monitors its operational health and ensures its capacity to meet consumers' needs~\cite{S4,S48,S110}. \textcolor{black}{The gray literature mentions several metrics for assessing the operational health of data products: data quality metrics, uptime, performance, and cost~\cite{S4,S48,S71}.}

\end{description}

\paragraph{Automation of Governance Activities.} Automation plays a significant role in the federated computational governance model to ensure it can effectively scale throughout an organization~\cite{S1,S2,dehghani2022data}. \textcolor{black}{According to gray literature sources~\cite{S57,S98,S110}, automation is key to coping with the data governance challenges introduced by frequent regulatory changes and increases in product consumers and data sources. They also provided examples for data governance actions that can be automated~\cite{S21,S57,S88,S98,S110}:} data classification (\eg sensitivity labels), data anonymization and encryption, data quality checks, data retention and deletion to ensure compliance, and automated monitoring, including alerts and automatic responses (\eg scale a resource to meet demand).   

\subsection{Principle 4: Self-Serve Data Platform}
\label{self-serve}
Data mesh advocates building domain-agnostic infrastructure and platform services and offering them in a self-service manner to empower different actors involved in creating, deploying, and maintaining data products~\cite{S1,S2,dehghani2022data}. The infrastructure services are to provision and manage computing resources such as VMs, networks, and storage disks. The platform services support the design, development, deployment, and management of data products, governance applications, and their components (\eg data pipelines, ML pipelines, and microservices). 

\paragraph{Objectives.} \textcolor{black}{According to Zhamak~\cite{S1,S2,dehghani2022data}, the self-serve platform aims to optimize the experience of data mesh stakeholders (\eg product developers, consumers, and governance teams) in their respective tasks, \ie developing, consuming, and governing data products. The gray literature is in agreement with Zhamak. First, the self-serve platform can reduce product developers' required level of specialization as they do not need in-depth knowledge of managing infrastructure and platform services~\cite{S4,S9,S25,S43,S72}. Second, it can increase the efficiency of developing and managing products as there is no duplication of efforts from each domain to manage platform services~\cite{S7,S9,S43,S53,S55,S65,S71,S92}. Last, the common platform services can improve the uniformity of data products, reduce friction between domains, and facilitate interoperability between data products~\cite{S19,S33,S88,S93}. }

\paragraph{Building Self-serve Data Platform.} A team of highly specialized infrastructure and platform service developers builds and maintains the platform~\cite{dehghani2022data}. This central platform team has to be in close contact with all domains to identify and prioritize the requirements for platform services, their features, and their interfaces~\cite{S85,S86}. \textcolor{black}{The gray literature recommends using IaC (Infrastructure as Code) blackprints to simplify the provision of data products using the self-serve platform~\cite{S7,S13,S53,S54,S93,S97,S98,S105,S108,S111}.} The IaC templates encompass predefined configurations for infrastructure resources and platform services and integrate security measures, policies, and standards in line with those set by the federated governance team. For example, the data platform team can develop IaC templates for different data product types (\eg data pipelines, ML pipelines, and ML model services) and different product component types (\eg a data ingestion component using the Kafka message broker, a data store component using Google object store, and a training pipeline executor component using the AWS SageMaker). The product developers can find the correct IaC templates, customize them as necessary, and use the updated templates to provision and manage their data products. 

\paragraph{Components of Self-Serve Data Platform.} \textcolor{black}{Zhamak proposes a logical decomposing of the platform into three layers: data infrastructure utility plane, data product experience plane, and mesh experience plane~\cite{S1,dehghani2022data}. They aim to encapsulate the capabilities to manage infrastructure and platform resources, build data products, and govern the data mesh. A few gray literature sources discussed this logical architecture of the self-serve platform~\cite{S99}. However, many practitioners provided examples for platform components. }  

\begin{description}
 
 \item[\textcolor{black}{Fundamental Computing Resources.}] \textcolor{black}{The platform should offer compute, network, and storage resources, including virtual machines, network devices, and object stores. The literature sources mostly mention using resources from public IaaS (Infrastructure as a Service) providers~\cite{S6,S13,S43,S78,S102,S103,S110}. } 

 \item[Polyglot Data Stores.] 
 
 \textcolor{black}{Regarding data storage, the common recommendation from the practitioners~\cite{S3,S7,S10,S31,S37,S53,S58,S80,S97,S98} and Zhamak~\cite{S2,dehghani2022data} is polyglot data stores that enable using multiple data storage technologies transparently within a single system.} Data products might contain different data types that require different data stores. For example, a data product might consist of highly structured, tabular data stored in a relational database. In contrast, other data products might consist of highly unstructured data that can best be stored in an object store (\eg videos or images). Thus, the platform must support the storage and serving of polyglot data.
 
 \item[Services for Data Product Components.] Data products generally consist of components such as data ingestion, data (ETL) pipeline, ML training pipeline, ML model storage, ML feature storage, message broker, and API~\cite{S7,S12,S31,S37,S42,S58,S64,S72,S73,S98,S108}. The product developers should be able to use platform services to implement and manage such product components. For example, the developers should be able to use the pipeline orchestration tools from the platform to develop, deploy, schedule, execute, and monitor data and ML pipelines in a self-service manner. 
 
 \item[\textcolor{black}{Metadata Repository/Data Catalog.}] Products, their components, and their artifacts (e.g., ML models and database tables) have various metadata; thus, the platform needs a service to generate, store, and manage metadata. Moreover, company-wide metadata, such as standardized definitions of business terms and reference data, should also be available so product developers can incorporate them into their data products~\cite{S90,S113,S114}. \textcolor{black}{As mentioned in \Cref{dataproducts}, a data catalog can be used to develop a product catalog service that stores product metadata to enable the discovery of data products.}

 \item[Federated Query Engine.] The self-serve platform should offer a federated query engine to query data from different domains and sources, \textcolor{black}{enabling product aggregators to create composite data products easily~\cite{S16,S31, S37,S98,S110}.}
 
 \item[Monitoring.] The platform should offer tools that enable domain teams to monitor their data products and the federated global governance team to monitor the data mesh as a whole~~\cite{dehghani2022data}. The tools should include but not be limited to functionality that enables monitoring of the operational health, lineage, costs, and performance concerning data quality metrics~\cite{S4,S48,S71}. They should also support logging, alerts, and automated interventions (e.g., scaling up resources)~\cite{S4,S33}. Finally, the monitoring tools must be integrated with the other relevant platform services, for example, adding monitoring capabilities to data pipelines and ML pipelines~\cite{S72}. 
 
 \item[Product Lifecycle Management.] The platform should offer capabilities (\eg product versioning, source control, and testing) to allow product developers to manage the lifecycle of a data product~\cite{S4,S53,S64,S110}. Developers should be able to apply DevOps~\cite{ebert2016devops} practices when deploying and operating data products~\cite{S27,S31}. For example, they should be able to use CI-CD (Continuous Integration and Continuous Delivery or Continuous Deployment) platform services to atomically build and test the data pipelines and APIs used by the products, deploy them using the pipeline orchestration and container hosting services of the self-serve platform.
 
 \item[Security and Privacy.] The platform services must enable the realization of data products' security and privacy requirements~\cite{S2,dehghani2022data}. For example, the platform can include a data encryption service to secure data at rest and in motion~\cite{S13,S28,S86}. In addition, an identity and access management (IAM) service can be provided to manage user and product identities and enforce access control policies for data products and other platform services~\cite{S7,S28,S29,S79,S103,S110}. 
 
 \item[Policy Enforcement.] The platform should offer tools that allow local and global federated governance teams to define, store, attach, observe, and enforce governance policies~\cite{S2,dehghani2022data}. \textcolor{black}{As mentioned in \Cref{dataproducts}, the gray literature recommends using the \textit{policy-as-code} approach for automating policy enforcement. Thus, the platform needs to support necessary tools such as OPA(Open Policy Agent) engines~\cite{S29,S78}.} Moreover, the platform must also provide services to support policy implementation, \eg data anonymization service and data quality metrics calculation service~\cite{S12,S45,S73,S97,S110}. 
 
 \item[Business Intelligence (BI) Tools.] \textcolor{black}{BI dashboards and reporting applications are common types of consumer-aligned data products~\cite{S10,S110,S112}.} Thus, the platform should enable business users to explore data, create visualizations, generate insights, and create reports in a self-service manner~\cite{S31,S64,S66,S98,S108}.
 
\end{description}

\section{RQ2. Motivations, Benefits, Concerns, and Applicability} 
\label{benefits}
\textcolor{black} {
This section presents the motivation for the data mesh approach, its perceived advantages and disadvantages, and the factors determining its adoption in organizations. 
}

\paragraph{\textcolor{black} {Motivations.}} \textcolor{black} {Data Mesh was developed as a direct response to centralized monolithic data management approaches that have been industry standards, such as data warehouses and data lakes. Based on the literature examined in this survey, we find three factors that motivate organizations to adopt a data mesh paradigm in their organization.}

\begin{description}
 \item[\textcolor{black} {Scaling with Data Sources.}] \textcolor{black} {As the number of data sources grows within an organization, it becomes increasingly difficult to maintain an overview of the various sources and guarantee their high-quality delivery. This problem became especially apparent during the rise of big data and has been one of the main motivations for developing data lakes in favor of data warehouses~\cite{Giebler2019}. Despite the widespread adoption of data lakes, gray literature demonstrates organizations still struggle to discover and make available all data within the organization ~\cite{S17,S23,S53,S48,S52,S58}, make various data sources interoperable ~\cite{S23,S35,S78,S48} and establish data quality and consistency at the source~\cite{S3,S23,S88,S35,S77}. Data mesh proposes to solve these issues at the source rather than address them centrally.}

 \item[\textcolor{black} {Central Teams as a Bottleneck.}] \textcolor{black} {The flip side to the ever-increasing number of data sources is the growing number of data-driven use cases in organizations. Usually, the data for these use cases is provided by a centralized team of hyperspecialized data engineers who also manage the monolithic data platforms in which the data is aggregated. However, relying on centralized teams hinders scalability and agility in data delivery, which is required to address the diverse needs of various data consumers \cite{S3,S5,S35,S42,S43,S4,S56,S60,S61,S77,S38,S88,S100, S110}.}

 \item[\textcolor{black} {Separation of Provider and Consumer Expertise.}] \textcolor{black} {Another problem, which the previous problem compounds, is that, in most organizations, data provider expertise is separated from data consumer expertise. Effective data usage requires a deep understanding of the data and the proposed use case. Relying on centralized teams to understand both the context of the providing domain and the consuming domain puts additional stress on the existing bottleneck, whereas the data provider and data consumer can be counted on to know their respective domains ~\cite{S9, S16, S13, S43}.}

\end{description}

\paragraph{Benefits.} From the gray literature, we identified the following potential benefits of data mesh compared to traditional data architectures. We briefly discuss each identified potential benefit's relation to the problems that drive data mesh transition, as identified above.

\begin{description}
 \item[Scalability.] A data mesh architecture is much more scalable than a centralized data architecture since dependencies on a central management platform team are reduced, which allows the organization to scale horizontally (\ie more data sources) and vertically (\ie more consumers) without creating a bottleneck~\cite{S3,S4,S12,S13,S18,S25,S28,S48,S60,S62,S6}. 
 
 \item[Increased Agility.] The benefits mentioned above make organizations more agile as they reduce the reliance on centralized teams and bring data providers directly in contact with data producers. By removing the bottleneck, they can deliver new data applications at a faster rate and can thus respond quickly to changing needs and goals~\cite{S3,S25,S28,S34,S40}. Overall, organizations can create more data-driven applications that unlock the value of their data~\cite{S92}.
 
 \item[Higher Data Quality.] By making those close to the origin of the data responsible for offering it as a data product, accountability for data quality is established~\cite{S10,S13,S17,S37,S60,S87,S108,S109}. Domain teams can use their domain-specific knowledge to build data products that are valuable to various customers~\cite{S7,S23,S42,S69}. The federated governance team defines a set of global standards and policies that the domains must adhere to when creating data products~\cite{S3,S9,S25}.
 
 \item[Better Data Discovery] In the data mesh, domain teams are incentivized to maximize the usage of domain data products and, hence, to make the products easily discoverable~\cite{S40,S50,S60}. Therefore, they create self-describing data products and publish product metadata to the central product registry/catalog for easy discoverability~\cite{S10,S87}. Improving the discoverability of data (products) lets organizations effectively scale the amount of data in their ecosystem.
 
 \item[Better Governance.] Data mesh also allows for better big data governance. The decentralized approach is more suitable for governing a high volume of data by distributing governance responsibilities to the domains close to the origin of the data~\cite{S3,S12,S10}. Because data providers are in direct contact with data consumers, they can better judge who should and should not be able to use certain data in certain conditions and facilitate organizations to comply with external regulations~\cite{S28,S60,S95}.

 \item [Reduced Data Lead Time.] Lastly, domains can independently provision and maintain data products, and consumers can independently use these products through the self-service model without needing to rely on any centralized team. Moreover, domains and federated governance teams are empowered through the self-service platform, dramatically reducing the data lead time throughout the organization~\cite{S3,S53,S69,S77,S82}.
\end{description}

\paragraph{Concerns and Disadvantages.} Data mesh comes with its own set of challenges. Below, we present an overview of the main concerns for the data mesh recognized by the gray literature. 

\begin{description}
 \item[Change management.] Moving to a data mesh architecture will challenge organizations from technical and organizational perspectives~\cite{S3,S25,S16}. Organizations already have established data management processes and a hierarchical organizational structure that will likely need restructured~\cite{S16,S28,S35,S87}. \textcolor{black}{The employees may be intimidated by changes and resist. To overcome this challenge, the organizations can train their teams on the cultural shift (the way of working)~\cite{S35,S63}, educate the stakeholders about the bigger picture and the overall benefits of the transition to the data mesh~\cite{S58,S63,S87}, and empower the existing business domain teams with data engineers~\cite{S63}. The organizations can create an enabling team to work closely with domain teams to help onboard and consume data, train, and guide teams on data mesh best practices~\cite{S109,S110}.} 
 
 \item[Lack of Talent.] Data mesh requires domain teams to start operating independently, so the teams need members with the necessary technical skills (\eg data engineering and machine learning). While a vital goal of the self-serve platform is to empower the developers of less technical expertise to develop data products, building self-serve platform services can be challenging and require time and effort for research and development. \textcolor{black}{Unfortunately, there is a lack of experienced data engineers~\cite{S16,S41,S46,S48,S77}, which was also a common bottleneck in the data warehouse and lake architectures~\cite{S77}.}  

 \item[Data Duplication.] As domains repurpose data for new use cases in composite data products, data from source domains may be copied and duplicated. In particular, when implemented on a large scale, data mesh can potentially increase data management costs~\cite{S3,S18,S24,S25,S28,S109}. Furthermore, as data is stored in different organizational locations, it will become more challenging to govern data~~\cite{S41,S48,S109}. \textcolor{black}{The remedies proposed by the practitioners include creating data virtualization techniques (e.g., zero-copy data access) that enable ingesting data without copying them and developing open protocols for sharing data securely and efficiently~\cite{S35,S41,S44}. }
 
 \item[Effort Duplication.] Even though the self-serve platform tries to reduce the duplication of efforts concerning data platform management, there will still be a certain degree of duplication of efforts and skills for building and maintaining pipelines throughout the different domains compared to a team working on centralized data~\cite{S4,S9,S18,S35,S85}.

 \item[\textcolor{black} {Technical Debt.}] \textcolor{black} {Multiple data products and pipelines developed by independent teams may not adhere to a common set of software development best practices and principles~\cite{S3, S21, S25}. Using improper practices when developing data products can lead to a huge technical debt. Hence, a mature federated governance model can reduce the technical debt by accurately recognizing and enforcing standards, best practices, and so on~\cite{S3,S21,dehghani2022data}.}

 \item[\textcolor{black} {Complexity.}] \textcolor{black} {The distributed and decentralized nature of the data mesh increases the overall complexity as it requires managing and coordinating many independent and diverse units~\cite{S29,S97,S109}. The data mesh can also introduce more interoperability challenges for organizations that adopt a multi-cloud approach~\cite{S35}. The complexity can decrease as the self-service platform and computational governance model mature~\cite{dehghani2022data}.} 
 
\end{description}

\paragraph{Applicability.} \textcolor{black} { The gray literature recognizes several conditions for which the data mesh is more suitable than a centralized data architecture.}

 \begin{description}
 \item[Large and Diverse Data Landscape.] Organizations with a large data landscape and many data providers and consumers might benefit from a decentralized architecture~\cite{S3,S4,S5,S7,S8,S22,S25,S26,S58,S62,S68}. Also, when organizations possess diverse data sets that frequently change over time, the data mesh might serve better than a centralized approach with many point-to-point connections~\cite{S9,S22,S26,S97}.

 \item[Need for Agility.] Organizations whose data platform and central data provisioning teams reduce their innovative capabilities and require a faster time to market for data applications can benefit from a data mesh architecture~\cite{S3,S4,S5,S12,S22,S35}. Also, when there is a need to experiment with data frequently, data mesh is more beneficial than a centralized architecture as it removes the dependency on an oversized and overloaded data provisioning team~\cite{S4,S61}.

 \item[Need for better Data Governance.]
 Centralized data lakes can quickly become data swamps without data ownership and governance models. Data mesh can resolve these issues for organizations needing clear data ownership and highly governed data based on domain knowledge~\cite{S7,S9,S58}. In addition, organizations that face many external regulations can use data mesh to ensure compliance as the data mesh facilitates better data governance~\cite{S13,S28,S76}.  
\end{description}

\textcolor{black} {Data mesh may not be the most practical data architecture for all organizations. The gray literature mentions several scenarios where the data mesh approach may not be the most suitable option.}  

 \begin{description}
 \item[\textcolor{black} {Lack of Domain-Orientation.}] \textcolor{black} {The data mesh presumes an organizational structure of independent cross-functional teams aligned with the structures of business domains. If an organization (e.g., typically small and medium-sized companies) does not follow domain-driven design principles and does not have multiple teams who can make data products independently, then the data mesh approach may not fit the organization~\cite{S5,S17,S26,S35,S109,S110}. }

 \item [\textcolor{black} {Low Data Governance Maturity Level.}] \textcolor{black} {The data mesh needs matured data governance support to prevent data silos and to ensure the regulatory compliance of data processing across autonomous domains. Thus, it may not be suitable for organizations with a low level of maturity in data governance (e.g., no automated data quality assessment or regulatory compliance checking)~\cite{S12,S26,S63,S109,S110}.}

 \item [\textcolor{black} {Low Data Platform Maturity Level.}] \textcolor{black} {The self-serve data platform is key to reducing the cost of owning data products and implementing data governance across domains. Building a self-serve data platform can be impractical for organizations without experienced data platform engineers~\cite{S26,S110}.} 

 \item [\textcolor{black} {Lack of Data Use Cases.}] \textcolor{black} {When an organization has a small number of useful analytical data sources and a few data use cases, the cost and risks of transitioning a data mesh would far outweigh the benefits~\cite{S12,S26,S110}.} 

\end{description}
\section{RQ3. How should an Organization Build a Data Mesh?}
\label{referencearchitecture}
This section presents a set of reference architectures to describe the data mesh architecture in detail. We leveraged the relevant architectures used in the service-oriented computing domain and the insights from our gray literature review.

\subsection{A Layered Architecture of Capabilities and Roles in Data Mesh}
\label{lacap-ra}
\begin{figure}[ht!]
 \centering
 \includegraphics[width=\textwidth]{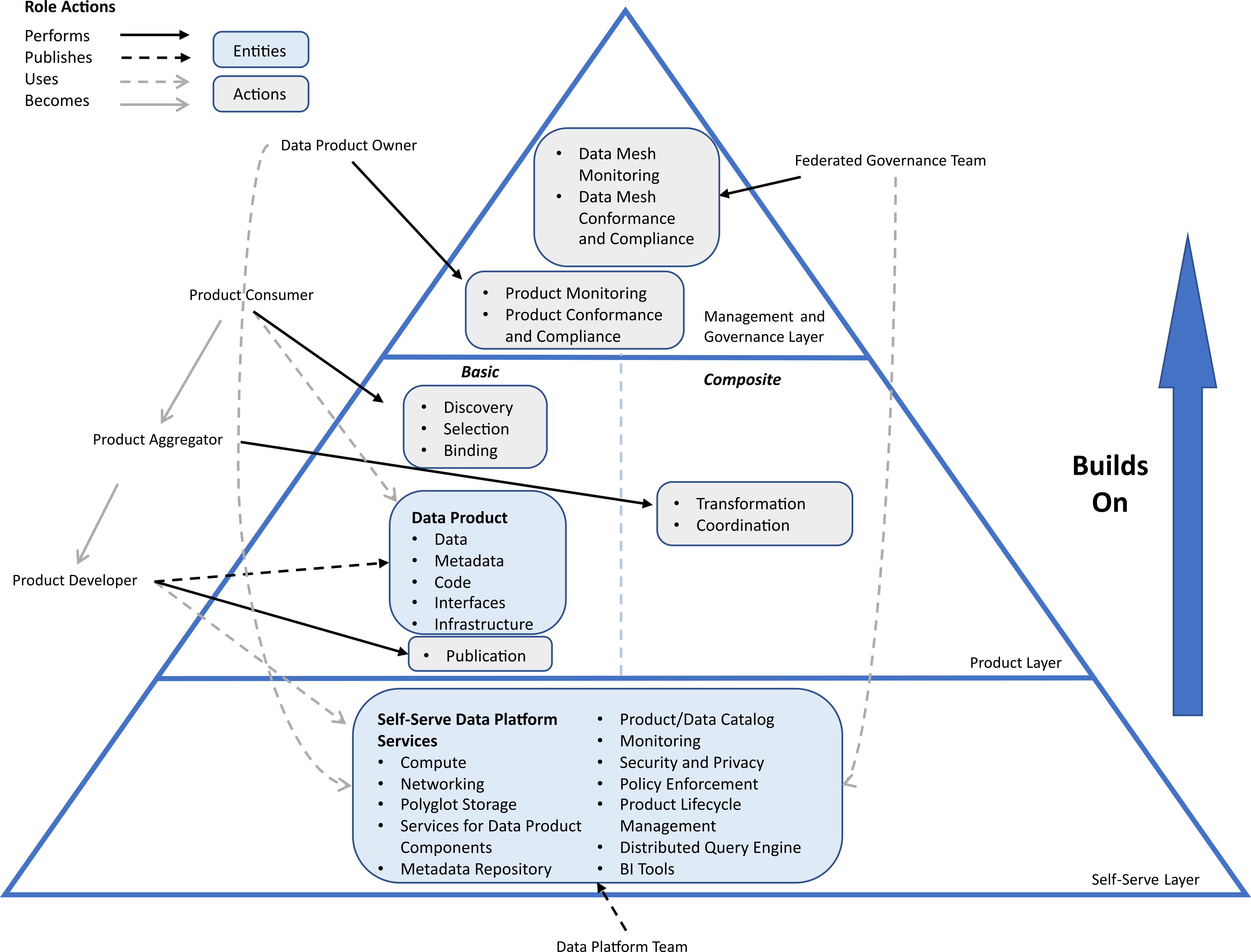}
 \Description{The layered architecture of capabilities and roles in data mesh.}
 \caption{The layered architecture of capabilities and roles in data mesh.}
 \label{fig:layered-architecture}
\end{figure}

This section introduces a layered model to logically organize different constructs, roles, and responsibilities in a data mesh architecture. We mapped the relevant findings from our gray literature study into the xSOA model~\cite{papazoglou2007service,Papazoglou2}, which aims to group and logically structure the capabilities of complex SOA applications. \Cref{fig:layered-architecture} shows the layered model of the data mesh. The separation into layers highlights the different needs regarding i) providing the infrastructure and platform services necessary for realizing and consuming data products, ii) creating, consuming, and composing data products, and iii) managing data products throughout their lifecycle. It is important to highlight that although higher layers build on lower layers, they only represent a separation of concerns and do not imply a hierarchy. 

\textcolor{black}{
As discussed in \Cref{dataproducts}, Zhamak and the gray literature categorize data products into source-aligned, aggregates, and consumer-aligned. By adopting service types in SOA, we classify them into \textit{atomic} and \textit{composite} data products. Like composite web services, composite data products use other data products (atomic or composite) for their source data and integrate them to create new value-added products. Hence, aggregates and consumer-aligned products are composite products. The source-aligned data products are atomic because they only ingest data from the operational source systems.}

The layered architecture shown in \Cref{fig:layered-architecture} does not depict all responsibilities in a data mesh architecture. Instead, it provides a view of the responsibilities needed to create individually managed data products. For example, the federated governance team’s responsibility to create a business glossary is not included in this architecture because this is not related to building a specific managed data product at runtime.

\paragraph{Roles.} We identified six different roles for data mesh actors:

\begin{description}
 \item[Data Platform Teams] are responsible for building and maintaining the self-serve data platform~\cite{S2,S22,S42,S73,S85,S86}. They also create the IaC blackprints that product developers can use to provision and configure infrastructure resources and platform services for hosting and managing data products. \textcolor{black}{They need to possess software and data engineering skills~\cite{S22}.}
 \item[Product Owners] are responsible for offering and governing the data products in their domains~\cite{S1,S2,S3,S16,S18,S25,S33,S35,S43,S49,S67,S73,S82,S86,S81,S88}. They must ensure that the data products are interoperable and meet the requirements of the consumers. \textcolor{black}{They are generally responsible for domain-level governance activities (see \Cref{governance}).} 
 \item[Product Developers] build and publish data products on the data mesh using the tools/services the self-serve platform provides~\cite{S1,S2,S16,S18,S21,S35,S39}.
 \item[Product Aggregators] build and provide composite data products. The gray literature does not explicitly mention the aggregator role. We introduced it to clearly distinguish between the responsibilities and capabilities of building source-aligned atomic data products from composite data products. 
 \item[Product Consumers] use data from the data products~\cite{S10,S13,S19,S54,S91}. Product aggregators are also product consumers since they consume multiple other data products and aggregate them into composite data products.
 \item[Federated Governance Teams] comprise representatives from different domains across the organization and govern the mesh globally~\cite{S1,S16,S21,S25,S28,S110}. \textcolor{black}{\Cref{governance} discusses the governance activities of the federated governance team.}
\end{description}

\paragraph{Self-Serve Layer.}
\label{la-self-serve}
Self-serve platforms offer a set of tools/services in a self-service fashion that the actors involved in a data mesh architecture can use to fulfill their responsibilities (see \Cref{self-serve}). For example, product developers can use the platform services to develop, test, deploy, and evolve their data products. In contrast, the product owners and the federated governance team can use platform services to monitor and govern individual data products and the overall data mesh. 

\paragraph{Data Product Layer.}
\label{la-dataproduct}
The data product layer sits on top of the self-serve layer. As discussed in \Cref{dataproducts}, data products consist of data, metadata, code, interfaces, and the infrastructure to deploy the product. Once the product developers define and implement all components, the data products can be published in a product catalog. Product consumers browse the catalog to discover candidate data products for their use cases and check the corresponding metadata to verify that they suit their needs. Product metadata can also enable consumers to automatically connect (binding) to the data products to obtain their data. \textcolor{black}{For example, if a data product uses a storage bucket as a serving endpoint, it can publish the metadata, such as the URL of the bucket and the names and schemas of the result objects, to the catalog~\cite{S45}. }

Composite data products use data from other (atomic or composite) data product(s), potentially apply complex transformations, and publish the results as new data products (see \Cref{dataproducts}). Composite data products contain the same components as atomic data products. However, product aggregators perform additional actions to build composite data products, similar to the additional responsibilities in xSOA~\cite{papazoglou2007service}. To distinguish between these responsibilities, the layered architecture in \Cref{fig:layered-architecture} differentiates between product developers, who build atomic data products, and product aggregators, who create composite data products. 

Composite data products need to provide new value to the organization. A simple integration of two data sources can easily be replicated and may not provide sufficient additional value. Therefore, a product aggregator may need to perform complex transformations on the data from upstream data products to create value-added data. Furthermore, composing products may involve coordinating a series of steps,\eg, ingesting data from multiple data products, merging the data from a subset set of products, cleaning and standardizing data, and validating and storing intermediate and final results. 

Composite data products must conform with the upstream data products throughout their lifetime. For example, upstream product developers might change their data products (\eg schema and product interfaces), which can affect downstream consumers~\cite{S53,S55,S72,S110,S110}. Product aggregators must deal with such issues by adopting suitable strategies such as product versioning and using an anti-corruption layer (an adapter layer)~\cite{S72,S110}. 

\paragraph{Management and Governance Layer.}
\label{la-management}
Data products must be managed and governed at runtime to ensure they provide value to the organization. As discussed in \Cref{governance}, there are two levels of governance: local and global. The local governance (per each product) is responsible for the quality and conformance of individual data products. For example, the local governance team must ensure that a data product operates within the specified SLOs throughout its lifetime and remains compliant with local and global policies and external regulations. The owners of the composite products also need to monitor the performance of upstream data products in terms of service level objectives (SLOs). The performance of the composite data products can be affected by that of upstream data products. Thus, a product aggregator needs to monitor the performance of upstream products to determine any SLO violations and ensure the composite products operate within its SLOs. The main goal of global governance at runtime is to ensure that the mesh architecture can scale throughout the organization effectively. The federated governance team monitors the mesh to assess the data mesh's overall health and enforce the global policies, interoperability standards, and guidelines. 

\subsection{A Layered Architecture for Data Mesh Development}
\label{la-ra}
\begin{figure}[ht!]
 \centering
 \includegraphics[scale=0.80]{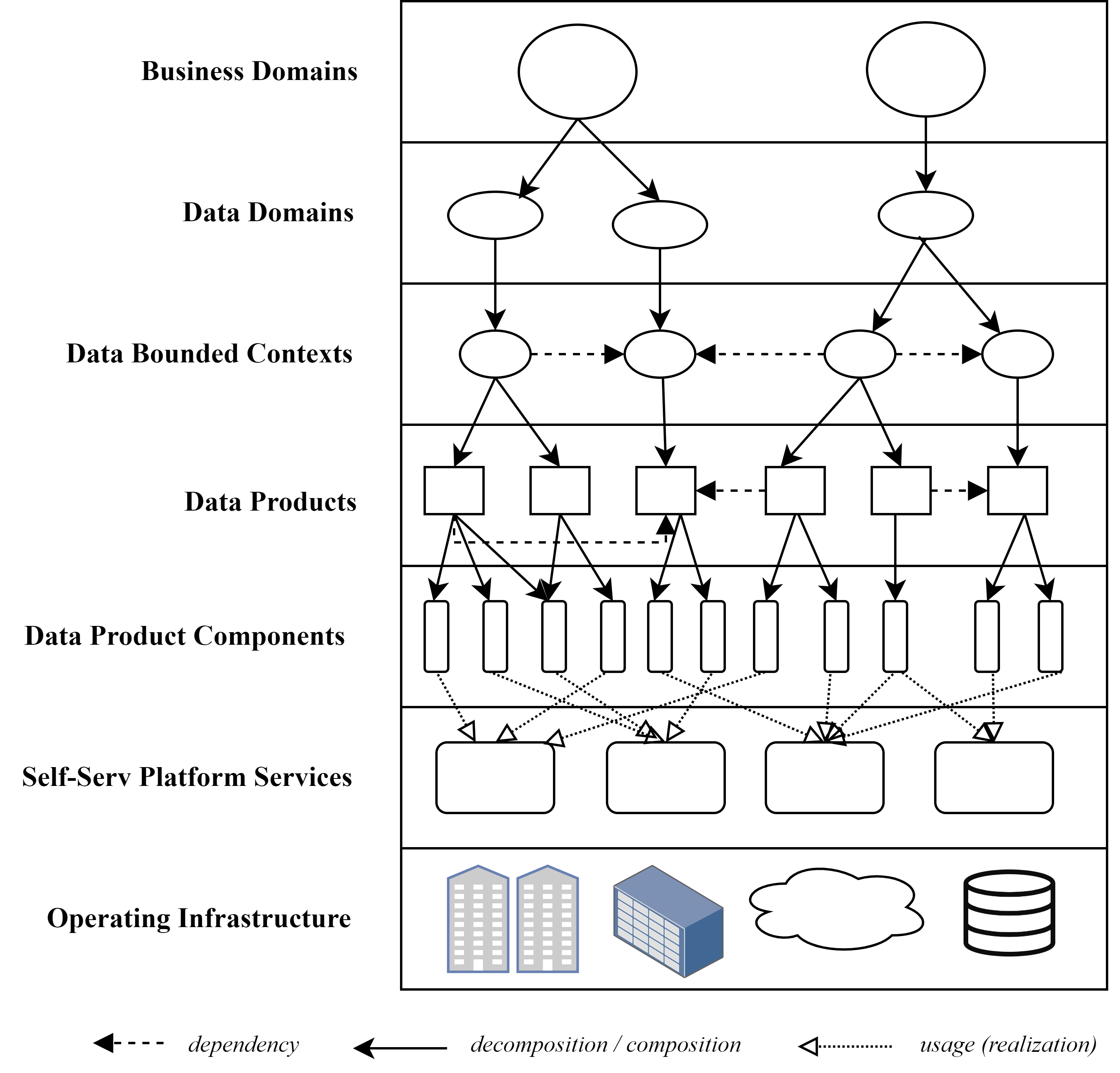}
 \Description{A layered model for the data mesh development.}
 \caption{A layered model for the data mesh development.}
 \label{fig:decompositiondomains}
\end{figure}

The data mesh literature does not explicitly provide an end-to-end methodology for implementing data mesh in or across organizations. However, the literature advocates applying the principles of domain-driven design (DDD), product thinking, and platform-based development. In particular, data products should be identified by applying DDD to the data produced by the business domains of the organization~\cite{S1,S2,S11,S36,S37,S39,S45,S46,S53,S55,S58,S71,S75,S78,S79,S87,S90,S91,S96,S98,S103,S110,S111,S112}. Inspired by a layered model for SOA development~\cite{papazoglou2008web}, we introduce a layered model for data mesh development that integrates the previously mentioned principles. \Cref{fig:decompositiondomains} shows the proposed layered approach to data mesh development. The model comprises seven distinct layers: business domains, data domains, data-bounded contexts, data products, data product components, self-serve platform services, and operating infrastructure.

Following the DDD method, the first layer decomposes an organization's business into disjoint domains. A business domain consists of a set of business processes that share a common set of business capabilities and collaborate to accomplish the business goals and objectives of the domain~\cite{papazoglou2008web}. These processes produce valuable data for the domain and the rest of the organization and can thus be offered as data products. For example, a manufacturing organization may include domains such as distribution, manufacturing, finance, and human resources~\cite{papazoglou2008web}. The distribution domain may comprise key business processes such as consumer purchasing, order management, and inventory. The purchasing process may produce data containing information about products or services, time of purchase, and the amount spent. This data may be exploited to build data products to answer queries about purchase history, customer buying patterns, and other relevant details, such as stock availability and product appearance. 

The second layer partitions the analytical data of a business domain into one or more data domains. Data domains define boundaries around the analytical data. From the viewpoints of data mesh and data governance, data domains categorize and organize data to assign accountability and responsibility for the data. As discussed above, different business processes in a business domain may produce multiple data. Due to the diversity and complexity of processes, the responsibility for each process may be assigned to a separate team~\cite{danilova2019process}. Similarly, the analytical data of processes may be owned by different teams, originating multiple data domains from the same business domain. The gray literature also recommends aligning business and data domains and creating new domains for shared data as appropriate~\cite{S36,S37,S39,S46,S53,S55,S56,S71,S75,S78,S87,S111,S112}. 

The third layer identifies the bounded contexts within each data domain. The gray literature recommends using bounded contexts for scoping data products~\cite{S1,S35,S45,S79,S87,S90,S98,S112}.  DDD designs and implements software systems based on domain models, which are the object models of the domains that incorporate both behavior and data in the domain~\cite{fowler2012patterns,evans2004domain}. Creating and maintaining a unified model for larger and more complex domains, where multiple teams work on separate subsystems, may not be feasible or cost-effective~\cite{fowler2012patterns,evans2004domain}. To tackle this issue, DDD partitions the solution space of a domain into bounded contexts, where each can have a separate domain model. A data domain can also be complex and include multiple logical groupings of data. Each group can have a separate data owner and model (\ie bounded contexts for decomposing data domains). For example, there may be different categories of customers: online only, physical only, third party, prospect, individual, group, corporation, government, and charity. Managing data for each or subset of these categories might differ, and multiple teams or individuals might be responsible for it. Consequently, multiple bounded contexts emerge within a customer data domain. 

Various dependencies (design time and runtime) may exist between data bounded contexts. The gray literature recommends using the strategic design patterns of DDD (\eg customer-supplier and anti-corruption layer) to integrate bounded contexts (\ie context mapping)~\cite{S111}. For example, a bounded context of a composite product can use the anti-corruption layer to minimize the impacts of the changes to the data models of the data products from the upstream bounded contexts~\cite{S110}. \textcolor{black}{
For example, a data product for sales funnel analysis uses the customer search data product to get customers' browsing history. The funnel analysis product can implement an anti-corruption layer to prevent the adverse effect of changes in browsing history data format.}

The fourth layer identifies data products within each bounded context. For building applications for a bounded context, DDD recommends applying so-called tactical design patterns: entity, value object, aggregate, service, repository, factory, event, and module. These patterns can be used to enrich a domain model of a bounded context, and the enriched model can be used to derive the application architecture components. For example, in a microservice architecture, services, APIs, and events can be deduced from the domain model using the tactical design patterns~\cite{9426760}. The data mesh gray literature~\cite{S71,S96,S103,S110} also recommends applying these patterns, for example, using domain main events for ingesting and serving data and implementing a log of data changes, and using entities and aggregates for identifying products and creating database schemas and product APIs. 

The fifth layer consists of the components that implement functional and non-functional requirements of data products. Different types of data products exist, such as datasets (offered as database views or APIs), data pipelines, ML training pipelines, and ML prediction services. Different data products may use different architectures. For example, data pipeline products may use popular data architecture styles such as Lambda and Kappa~\cite{Davoudian2020}, and ML prediction services may use architecture styles as model-as-a-service or model-as-a-dependency~\cite{9779697}. The resulting data products can have components like streaming data ingestion, ETL workflows, storage of raw, cleaned, and enriched data, model and feature storage, APIs, data quality assessors, and report generators. Data products can also share the implementation of some components, such as a generic data validating service or model validation service. 

The sixth layer consists of the self-serve platform services that support the end-to-end lifecycle of data products and their components, from design, implementation, testing, deployment, execution, and monitoring to evolution. For example, data product developers should be able to use an ETL platform service in a self-service manner to create ETL pipelines for their data products, deploy and execute the created pipelines on-demand or according to a schedule, monitor the performance, resource usage, and errors of the executed pipelines, and make changes to the pipeline as necessary. \Cref{self-serve} lists common platform services we found from the gray literature. The platform team is responsible for self-serve platform services. They may develop platform services from scratch, reuse and customize third-party software, use the platform services offered by cloud providers, and modernize existing platform services by incorporating self-service capabilities. 

The seventh layer mainly includes the infrastructure resources used by the platform services and data products and the existing systems in the organization that act as data source systems. 

In a data mesh, management and governance applications are used by the federated governance team, in addition to data products, to monitor and control data products. \Cref{fig:decompositiondomains} does not explicitly include those applications. However, since those applications may use the monitoring data of the data mesh, developers may develop and offer them as data products. Furthermore, the management applications should also be able to use self-serve platform services.

\subsection{A Logical Runtime Structure of Data Mesh} 
\label{ra-ra}
\begin{figure}[ht]
 \centering
 \includegraphics[scale=0.50, angle=90]{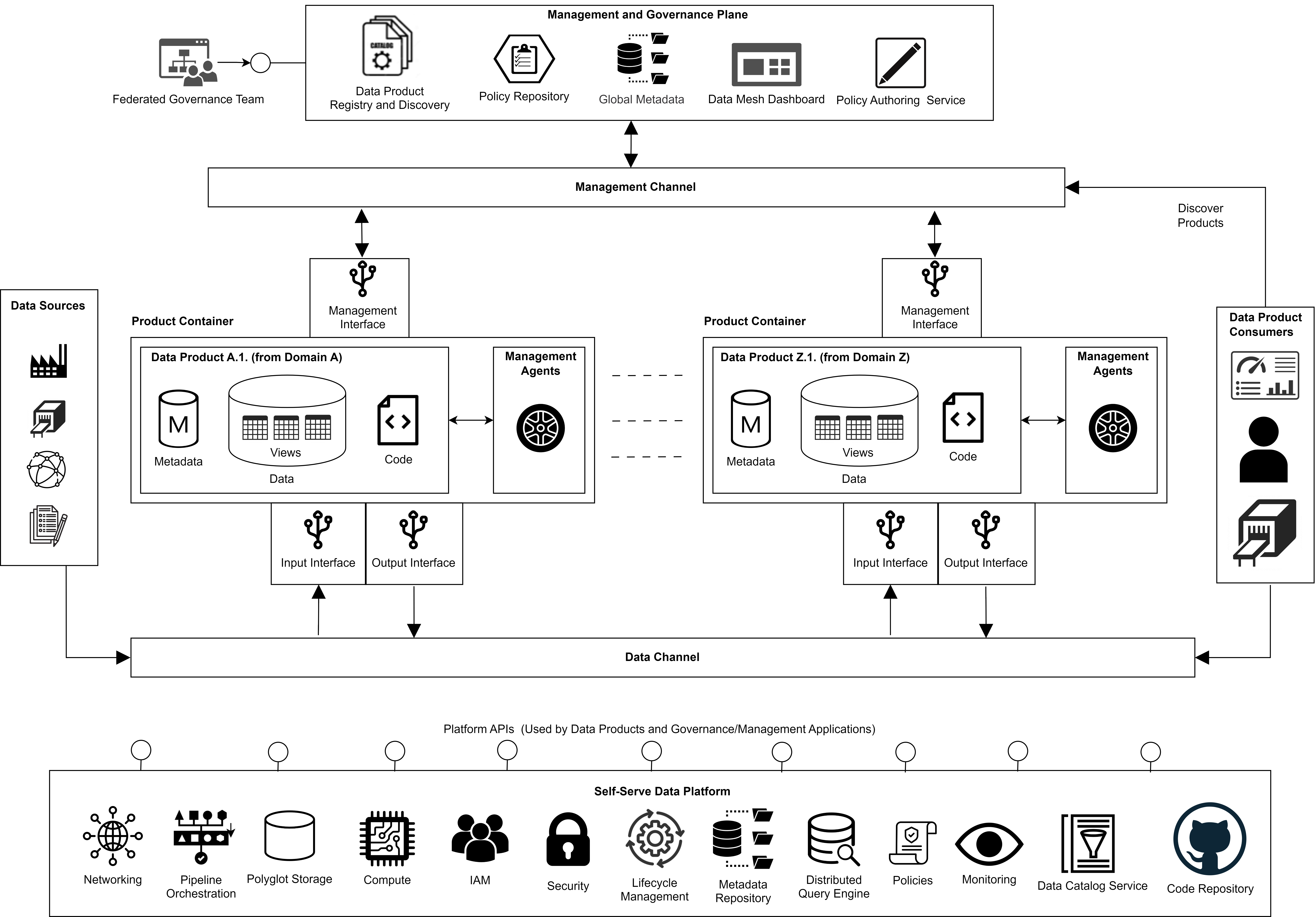}
 \Description{Logical runtime structure of a data mesh.}
 \caption{Logical runtime structure of a data mesh.}
 \label{fig:raruntime}
\end{figure}

This section presents a logical runtime view of a data mesh. The data mesh architectures presented in the gray literature source primarily focus on the runtime model of an inter-connected web of managed (governed) data products across data domains~\cite{S1,S8,S9,S10,S14,S29,S35,S42,S47,S53,
S58,S70,S71,S78,S80,S86,S99,S107,S110}. We extracted the essential components from those architecture designs and created a consolidated model by consulting web service management architectures~\cite{Papazoglou2,papazoglou2008web,servicemesh,SDSN}. \Cref{fig:raruntime} depicts the reference architecture at runtime. 

A product container is a deployment and runtime environment for a data product and its managing agents. We borrowed the concept of a container and a management agent from Web services containers and management agents~\cite{papazoglou2008web}. As in the web service management architectures~\cite{papazoglou2008web,Papazoglou2,servicemesh}, we separate the communication between data products, source systems, and end-users (data channel and input/output interfaces) and the communication between the management agents and the management/governance plane (management channel and management interface). The input interface~\cite{S51,S66,S72,S86,S97,S103,S108,S110} enables data products to receive data from source systems and other data products. The output interface~\cite{S51,S66,S72,S86,S97,S103,S108,S11} allows data products to provide their data to consumers or composite data products. Finally, the management interfaces~\cite{S86,S97,S103,S110} enable monitoring and controlling the behaviors of data products (\eg monitor product performance, raising alerts, controlling data routing, and enforcing security policies and data contracts). The modes of communication can be pull-based (\eg retrieving data via a REST API~\cite{S1,S64,S71,S75,S78,S82,S83,S90,S108,S111} or reading from a shared database/bucket~\cite{S13,S24,S86,S87,S89,S97,S98,S100,S112,S108}  or polling data from a message broker~\cite{S24,S29,S64,S78,S82,S83,S84,S86,S90,S101,S108,S110,S111}) or push-based (\eg pushing the data to the downstream consumers as the data become available~\cite{S13,S72,S101,S110,dehghani2022data}). 

The management agents~\cite{S2,S29,S78,S97} of a product implement management tasks at the product level, such as marking and enforcing policy decisions, collecting performance metrics, and managing the product life cycle (\ie creating, updating, and destroying products). The gray literature recommends using the sidecar design pattern and the policy-as-code approach to implement the monitoring and policy enforcement capabilities of the management agents~\cite{S29,S36,S78,S83,S113}. The sidecar pattern~\cite{burns2016design} decouples policy decision-making from the data product and can enable reuse and consistent implementation of monitoring and management logic across data products. The policy-as-code approach uses programming code to write policies, allowing applying software development best practices such as version control, automated testing, and automated deployment. A policy manifests as a set of rules that dictate the behaviors of the management agents, for example, how to enforce authentication and authorization policies for data product access, when to raise alerts, and where to route messages.

The components/applications in the management and governance plane can use the management interfaces of products to receive information (\eg performance data, access logs, and alerts) from data products and to send configuration and policy updates to the data products. We identified several key components for this plane from the gray literature: data product registry and discovery service~\cite{S4,S81,S88,S98}, policy authoring service and policy repository~\cite{S53,S110}, global metadata repository~\cite{S47,S53,S110,S114}, and data mesh dashboard~\cite{S53,S110}. The product registry and discovery service is connected to the management channel. Product developers publish the metadata of their products over the management channel. Consumers can discover data products by using the published metadata. \textcolor{black}{The global metadata repository can store metadata shared across domains, such as master and reference data.} The federated governance team can create global policies (\eg interoperability, security, and compliance) using the policy authoring services, store them in the policy repository for reuse and versioning, and enact them using the management interfaces provided by the individual products. The team can use the data mesh dashboard to monitor the operational health of the data mesh, including the cross-domain access and usage of data products and level of policy compliance. 

The self-serve platform plane provides infrastructure resources and platform services to develop, test, build, deploy, execute, monitor, and manage data products and governance components. For example, a data product can use the ML workflow platform service to create and run the training pipelines. The product registry and discovery service can use data/product catalog or metadata repository platform services. \Cref{self-serve} discusses the key platform services of a self-serve platform.

\section{RQ4. What are the Research Challenges concerning Data Mesh?}
\label{research-agenda}
In this section, we put forward a set of research challenges for data mesh that need to be addressed. We identified them by mapping the challenges mentioned by the practitioners in the gray literature to the relevant research issues from the service-oriented computing~\cite{Papazoglou2,WeiRe,papazoglou2007service} and data management~\cite{dustdar2012quality,stonebraker2018data,al2019systematic,JANSSEN2020101493}. 

\begin{description}
 \item[Standardizing Data Mesh.] The data products in a data mesh should be interoperable to enable the seamless composition of products and foster cross-domain collaboration. The data products also need to be portable so that self-serve platform services and governance applications can be modified or replaced with little or no modification to data products. The standardization of the data mesh can enable its interoperability and portability~\cite{S8,S35,S38,S61,S110,S111}. The standards are necessary for i) defining data products, their interfaces, and their composition; ii) data product publishing and discovery, quality assessment, deployment, and governance; iii) self-serve platform services and their interfaces; and iv) artifacts used/produced by data products (\eg data models, metadata, and ML models). The standardization efforts in the SOA and cloud domains can potentially help address these challenges. \textcolor{black}{For example, the Web Service Description Language (WSDL) supports modeling services with diverse endpoints, message types, and access policies (\eg HTTP endpoints, publish-subscribe topics, and FTP endpoints)~\cite{Papazoglou2}. Hence, WDSL and other related web service standards may help identify the concepts for modeling different types of data products. Topology and Orchestration Specification for Cloud Applications (TOSCA) is an OASIS standard language used to model the deployment topologies of complex applications~\cite{bellendorf2020specification}. It can be a suitable candidate for a technology-agnostic IaC language that can define the interoperable deployment blackprints for data products and self-serve platform components.}    
 
 \item[Methodologies and Tools for Data Mesh Development and Operation.] The data mesh literature~\cite{S1,S4,S28,S29,S35,S42,S98,S111} provides general principles for developing a data mesh architecture, \eg domain-driven design (DDD), product thinking, platform thinking, and computational governance. However, no tools and detailed (empirically validated) methodologies exist for applying these principles to develop and operate a data mesh. Organizations would also need guidance and tools for systematically migrating their legacy data architectures into data mesh while assessing costs and benefits. \textcolor{black}{The research issues may benefit from the rich knowledge about applying DDD for designing and implementing microservices~\cite{DDD4MS} and migrating legacy monolithic architectures into cloud-based microservices architectures~\cite{MSMigration}.}  
 
 \item[Data Product Life Cycle.] Data products are at the heart of a data mesh. Each phase of a data product life cycle (\eg identification, design, development, testing, deployment, discovery, composition, operation, monitoring, and evolution) can have open challenges~\cite{S1,S35,S45,S79,S87,S90,S94,S98,S112}. For example, the product identification phase decomposes the analytical data landscape of an organization into a set of data products. The partitioning process needs to support applying domain-driven design correctly, ensure the modularity and utility of data products, and consider reuse of the artifacts of the existing data architecture, such as data source systems, data pipelines, machine learning pipelines, documentation, and execution logs. Further exploratory research studies are needed to properly understand the research challenges in each phase of the product life cycle. Due to the comparability of a data product and a service, we believe the existing relevant research studies on SOA and microservices can help understand and address the above product challenges. \textcolor{black}{For example, there exist literature on web service discovery~\cite{Ngan2013} and composition~\cite{SHENG2014218}, DDD for microservices development~\cite{DDD4MS}, and metadata management techniques for data stores~\cite{MetadataMgt} and web services~\cite{TOSI201516}.}  
 
 \item[Self-serve Platform Services.] The domain teams need to be able to use domain-agnostic platform services in a self-service manner to easily and independently build and manage their data products. While there exist research studies on self-serve systems, e.g., self-service BI (business intelligence)~\cite{lennerholt2018implementation} and self-service portals for cloud applications~\cite{Mietzner}, they do not cover each platform service type, \eg data pipelines, ML pipelines, and IaC based data product provisioning. Moreover, the recommender systems can also be incorporated into platform services to assist users with building data products, \eg providing recommendations interactively when creating a data pipeline or IaC blackprint. Finally, since organizations may already use data platforms, they may need guidance and techniques to add self-service capabilities to their legacy data platforms~\cite{S2,S16,S33,S43,S53,S86,S108}. \textcolor{black}{The existing literature on recommender systems for software engineering investigates simplifying and assisting users in each phase of the software development lifecycle~\cite{SERec,ModelRE}. Many studies have explored generating code, including SQL and data processing code, from natural language descriptions~\cite{NLCodeGen}. LLMs (Large Langauge Models) have also shown great potential in this research area~\cite{gptse,zhang2024tree}.}
 
 \item[Data Mesh Governance.] While data governance has been an active research topic for several decades, as noted in several recent reviews~\cite{DG_ABRAHAM2019424,DG_DAVIDSON2023100454,JANSSEN2020101493}, there are still many open research issues. Data mesh, especially its principles of decentralized ownership of data and federated computational governance, may add new dimensions to data governance research issues. For example, the efficient composition of data products across domain boundaries (and potentially across organizational boundaries) depends on the ability of the federated governance mechanisms to ensure that data products are interoperable, comply with QoS (Quality of Service) and policy requirements, and in line with the company-wide data strategy~\cite{S1,S28,S35,S86,S111}. Moreover, the end-to-end data governance automation using the policy-as-code approach (i.e., computational governance) is largely unexplored. \textcolor{black}{The policy-as-code approach has been applied in use cases such as enforcing policies in microservice architectures (using the service mesh)~\cite{chandramouli2020building} and IaC-based deployment pipelines~\cite{policyiac}, as well as enforcing contracts in machine learning  environments~\cite{truong2021qoa4ml}.}   
 
 \item[Organizational Change Management.] As discussed in \Cref{lacap-ra}, data mesh introduces a set of new roles and responsibilities for teams and individuals in an organization. In particular, the responsibilities of domain data are shifted closer to decentralized autonomous domain teams. Hence, for organizations that use centralized operating models and organizational designs, embarking on a data mesh journey may require a significant organizational change (in addition to technological changes). Data mesh can also significantly impact an organization's existing data management and governance processes and practices. Thus, more studies are necessary to understand the organizational impacts of data mesh and the barriers to the adoption of data mesh, as well as to create guidelines and frameworks to perform and manage required organizational changes~\cite{S2,S16,S28,S33,S35,S61,S87}. \textcolor{black}{The research on migrating on-premise systems to the cloud and legacy applications to microservices has also studied organizational challenges and potential solutions for them~\cite{gholami2016cloud,MSOrg}. Further research could also help to understand and address the organizational challenges in data mesh transition.}  
\end{description}

\section{Related Work}
\label{background}
This section reports on the studies concerning data mesh and related surveys.

\subsection{Studies on Data Mesh}
Loukiala \etal describe migrating from centralized, monolithic data platform architectures to a decentralized data-mesh-like architecture at a large Nordic manufacturing country~\cite{migatraionstudy}. Their work emphasizes concrete steps that can be taken during such a migration and provides high-level contrasts between traditional data warehouse architecture and data mesh. 
Machado \etal present several works on data mesh anatomy and architecture. The first is an overview and explanation of data mesh concepts highlighted by two implementations in the industry (an online retailer, \ie Zalando, and a streaming platform, \ie Netflix)~\cite{10.1007/978-3-031-07481-3_2}. Then, in another paper, they introduce a ``domain model'' that contains the abstract classes of a data mesh and a high-level architecture~\cite{machado2021data}. Finally, based on these works, they investigate existing (commercial) tools that can be used to realize a data mesh implementation and validate their architectural model through a proof-of-concept implementation demo~\cite{MACHADO2022263}.
Similarly, Butte and Butte~\cite{Butte2022} present a more-complete reference architecture. However, it is unclear what their architecture is based on, as the paper provides no clear references to the sources. These are currently the only published academic works explicitly discussing data mesh architecture. However, their descriptions of the data mesh concepts lack a thorough exploration of the state-of-the-art in data mesh. Instead, they are almost exclusively based on the blog posts of Zhamak Dehghani~\cite{S1, S2}, as well as two corporate presentations: Zalando~\cite{zalando2020} and Netflix~\cite{netflix2020}.   

\textcolor{black}{
Kumar and Sujata~\cite{KumarSujata} describe the key concepts of the data mesh, present a template for a data processing pipeline within a domain, and provide an example of a data mesh deployment using AWS cloud.   
Li \etal~\cite{LiMesh} presents a case study of applying data mesh concepts to power system data management. They use a federated data processing engine and a metadata layer over raw data sources to build value-added data services. 
Podlesny \etal~\cite{podlesny2022cok}  discusses the potential data privacy issues that a distributed data mesh can have, focusing on exposing PII (Personal Identifiable Information) data through quasi-identifiers. Kraska \etal~\cite{BRAD} propose a visionary system that aims to provide a unified data processing interface over a federated set of data processing engines, allowing non-expert end-users to use the data to access different specialized engines.   
Dahdal \etal~\cite{Dahdal} presents an architecture for a middleware system on the edges that offers data-intensive applications for tactical battlefields as data products. The middleware offers common data management features such as data ingestion, storage, security, and APIs.
Hooshmand \etal~\cite{Hooshmand_Resch_Wischnewski_Patil_2022} proposes applying the data mesh approach to engineering product lifecycle management (PLM). They also suggest applying domain-driven design and using ontologies for data modeling.}

\subsection{Related Surveys}
Oliveira \etal investigated the state-of-the-art literature on `data ecosystems' in 2018 ~\cite{Oliveira2019}. Data mesh falls under their intended definition of data ecosystems, and their work discusses many elements that are relevant to data mesh. However, none of the studies investigated mention data mesh or any closely related term.
Another meta-study that investigates the activities and challenges in big data (eco)systems and their implications on architectures is presented by Davoudian and Liu ~\cite{Davoudian2020}. Again, the activities and challenges identified in this paper relate directly to those of data mesh. However, the literature discussed herein focuses entirely on traditional data lakes and data warehouses as state-of-the-art.
It is only recently that data mesh has started to appear in literature surveys. For example, Driessen \etal~\cite{Driessen2022} explicitly discuss data mesh, which they consider a special kind of (internal) data market. However, in this work, data mesh is considered only abstractly in the larger context of data-exchanging ecosystems, and no extensive discussion of data mesh architecture is provided.
Based on the existing literature, therefore, we conclude that there exists a knowledge gap on data mesh architecture: On the one hand, papers that explicitly discuss a data mesh architecture (\cite{migatraionstudy, MACHADO2022263}) do not consider a broad spectrum of sources on which to base their architecture definition. On the other hand, existing metastudies that look at the state-of-the-art in data ecosystems~\cite{Driessen2022, Oliveira2019,Davoudian2020} hardly discuss data mesh sources nor provide a reference architecture.

Given the work above, there are no reviews of the data mesh literature. Moreover, only a little academic literature is available on the topic, primarily based on a few gray articles. Therefore, this review paper systematically analyzed, assessed, and summarized such literature sources. 
\section{Threats to Validity}
\label{discussion}
This section discusses the potential threats to external, construct, and internal validity~\cite{wohlin2012experimentation} that may apply to our study. First, we note the risk of missing relevant studies for the systematic gray literature review. Since the field is developing rapidly, it can be interesting to repeat the gray literature review after some time to see how the topic matures as the industry adopts it more. Hence, we extended the study to include sources from September 2021 to May 2022, which yielded 35 new sources.

Additionally, as part of our methodology's article selection process, we rely on the authors' (subjective) expertise to determine a source's quality and relevance. Such a process risks introducing bias to the final selection. Therefore, two authors assessed each source, and an inter-reliability kappa coefficient was calculated to ensure bias was not too high.

Finally, the reference architecture was created systematically by examining various sources. However, as noted above, few complete, holistic architecture instances exist to compare with, and many sources only consider partial architectures. These partial architecture instances might lead to a sub-optimal holistic reference architecture. At the same time, we note that this problem is somewhat cyclical: a reference architecture can contribute significantly to the number of complete architecture instantiations, which in turn can improve future reference architectures.

\section{Conclusion and Future Work}
\label{conclusion}
Although enterprises are increasingly becoming data-driven, their centralized monolithic data architectures still prevent them from leveraging the full value of their analytical data. The data mesh is an emerging domain-driven decentralized architecture and sociotechnical paradigm that promises to address the limitations of centralized enterprise data architectures. Even though the data mesh is gaining significant attention in the industry, there is an evident scarcity of academic research on the topic. This paper systematically selects and reviews 114 industrial gray literature on data mesh to identify key design principles, architectural components, organizational roles, benefits, and concerns. The findings from the review were further enriched with the academic literature from similar domains to create reference architectures for its key dimensions. Finally, we identified research challenges that need to be addressed to realize the vision of the data mesh. We believe the identified open issues and reference architectures can provide a framework for future data mesh research.    

As for future work, we will propose domain-specific languages to define data products and decision-marking platforms for the three pillars of the data mesh: data as a product, self-serve platforms, and federated governance.
\textcolor{black}{We will create the metamodels for data products and contracts and develop a method to design and instantiate a data mesh architecture by leveraging the reference architectures provided in this paper.}
% ...

\bibliographystyle{ACM-Reference-Format}
\bibliography{main}
\end{document}